%% file: arxiv_v2.tex
\documentclass{article}
\usepackage[left=0.75in,right=0.75in,top=0.75in,bottom=0.75in]{geometry}
\usepackage{natbib}
\usepackage[colorlinks]{hyperref}
\hypersetup{bookmarks=true, citecolor=blue, urlcolor=blue, linkcolor=blue}

\usepackage{doi}

\usepackage[utf8]{inputenc}
\usepackage{amsmath}
\usepackage{amsfonts}
\usepackage{amssymb}
\usepackage{amsthm}
\usepackage{arydshln}
\usepackage{color, colortbl}
\definecolor{Gray}{gray}{0.95}

\let\emptyset\varnothing
\usepackage{optidef}
\usepackage{dirtytalk}
\usepackage{multirow}
\usepackage{float}
\usepackage{adjustbox}
\usepackage{comment}
\usepackage{array}
\usepackage{graphicx}
\usepackage{url}
\usepackage{array, makecell} %
\usepackage{longtable}

\usepackage{color}
\usepackage{pgfplots}
\usepackage{caption}
\usepackage{subcaption}
\usepackage{pdflscape}
\usepackage{booktabs} 
\usepackage{enumerate}
\usepackage{authblk}
\usepackage{blkarray}
\usepackage{bm}
\usepackage{marvosym} 
\usepackage{soul}
\usepackage{textcomp, xspace}
\newcommand{\smp}{\hspace{1.5pt}}

\usepackage{tikz}
\usepackage{varwidth}
\usetikzlibrary{shapes.geometric, arrows}
\usetikzlibrary{calc,arrows.meta,positioning}
\usetikzlibrary{arrows.meta}

\usepackage{algorithm}

\usepackage{lscape}

\makeatletter
\newenvironment{breakablealgorithm}
  {
     \refstepcounter{algorithm}
     \hrule height.8pt depth0pt \kern2pt
     \renewcommand{\caption}[2][\relax]{
       {\raggedright\textbf{\ALG@name~\thealgorithm} ##2\par}%
       \ifx\relax##1\relax 
         \addcontentsline{loa}{algorithm}{\protect\numberline{\thealgorithm}##2}%
       \else 
         \addcontentsline{loa}{algorithm}{\protect\numberline{\thealgorithm}##1}%
       \fi
       \kern2pt\hrule\kern2pt
     }
  }{
     \kern2pt\hrule\relax
  }
\makeatother

\makeatletter
\usepackage[noend]{algpseudocode}
\algnewcommand{\algorithmicand}{\textbf{\hspace{3pt} and \hspace{3pt}}}
\algnewcommand{\algorithmicor}{\textbf{ \hspace{3pt}or \hspace{3pt}}}
\algnewcommand{\OR}{\algorithmicor}
\algnewcommand{\AND}{\algorithmicand}

\usepgfplotslibrary{dateplot}

\definecolor{greenxllite}{RGB}{196,215,155}
\definecolor{bluexllite}{RGB}{149,179,215}
\definecolor{redxllite}{RGB}{218,150,148}
\definecolor{greenxl}{RGB}{155, 187, 89}
\definecolor{bluexl}{RGB}{79,129,189}
\definecolor{redxl}{RGB}{192, 80, 77}
\definecolor{OliveGreen}{RGB}{70,136,52}

\usetikzlibrary{patterns}
\allowdisplaybreaks

\newcommand{\Keywords}[1]{\par\noindent
{\small{\em \textbf{Keywords}\/}: #1}}

\title{Scalable Algorithms for Bicriterion Trip-Based Transit Routing}

\author[1] {Prateek Agarwal} 
\author[1,2] {Tarun Rambha}
\affil[1]{\small Department of Civil Engineering, Indian Institute of Science, Bangalore, India}
\affil[2]{\small Center for infrastructure, Sustainable Transportation and Urban Planning (C\textit{i}STUP), Indian Institute of Science,
Bangalore, India}

\date{ }
\input{figures/symbol_list}

\begin{document}
\maketitle
\vspace{-5mm}
\begin{abstract}
This paper proposes multiple extensions to the popular bicriterion transit routing approach---Trip-Based Transit Routing (TBTR). Specifically, building on the premise of the HypRAPTOR algorithm, we first extend TBTR to its partitioning variant---HypTBTR. However, the improvement in query times of HyTBTR over TBTR comes at the cost of increased preprocessing. To counter this issue, two new techniques are proposed---a One-To-Many variant of TBTR and multilevel partitioning. Our One-To-Many algorithm can rapidly solve profile queries, which not only reduces the preprocessing time for HypTBTR, but can also aid other popular approaches such as HypRAPTOR. Next, we integrate a multilevel graph partitioning paradigm in HypTBTR and HypRAPTOR to reduce the fill-in computations. The efficacy of the proposed algorithms is extensively tested on real-world large-scale datasets. Additional analysis studying the effect of hypergraph partitioning tools (hMETIS, KaHyPar, and an integer program) along with different weighting schemes is also presented. 

\vspace{3mm}
\Keywords{transit routing; shortest paths; multi-criteria optimization, hypergraph partitioning}
\end{abstract}

\section{Introduction}
\label{sec:Intro}
Finding the \say{shortest/best} path efficiently is a widely researched problem in network science. Particularly, in transportation networks, the problem of route planning can be divided into two categories---Personal Mobility Routing and Public Transit Routing (PTR). Considerable advances have been made in the past few decades in both categories, see  \cite{delling2009engineering} and \cite{bast2016route} for details. While the goal in both cases is similar, i.e., to find the \say{best} path between a given source and destination, solution methods for PTR differ primarily due to its inherent time-dependent nature as buses/trains arrive and leave periodically on fixed routes. Further, transit users have greater sensitivity to objectives other than travel time such as transfers, wait and walk time, cost, crowding, and comfort. This makes the problem challenging since users look for journeys that weigh the trade-offs between several attributes. At a broad level, the problem of journey planning in PTR can be divided into the following categories:
\begin{enumerate}
    \item \textbf{Earliest Arrival Problem:} Given a source, destination, and departure time, this problem aims to find a journey that reaches the destination as early as possible. 
    \item \textbf{Multi-criterion Problem:} For a source-destination pair and a departure time, the goal is to find a journey (or a set of journeys) based on multiple optimization criteria. The output of this problem is generally a Pareto-optimal set of journeys. Among the possible objective function combinations, the bicriterion problem with travel time and transfers is the most explored setting.
    \item \textbf{Range Problem:} In this problem, instead of a single departure time, we seek all optimal journeys departing within a specific time range (e.g., journeys departing between 0700--0730). If the time range is 24 hours, it is also referred to as a \textit{Profile Query}. 
\end{enumerate}

A routing algorithm generally solves one or more of the above-mentioned problems. Early approaches in this direction involved modeling a transit network as either a Time-Expanded (TE) graph \citep{muller2007finding, pyrga2008efficient} or a Time-Dependent (TD) graph \citep{cooke1966shortest, ziliaskopoulos2000intermodal} and running a variant of the Dijkstra's method \citep{dijkstra1959note}. \cite{muller2007timetable} and \cite{pyrga2008efficient} compared both approaches and showed that while the TD model results in a compact graph and better performance, the TE approach allows modeling more realistic situations such as transfers. Although several heuristics that accelerate shortest path computations have been proposed over the years \citep{dreyfus1969appraisal, brodal2004time}, their biggest drawback has been that they cannot fully account for the dynamic nature of transit networks. For instance, speed-up techniques such as graph contraction, destination pruning, bi-directional search, and precomputing labels work well for routing in road networks, but fail to give satisfactory results for PTR \citep{bast2009car}. With improvements in communication technologies and increased use of smartphones, travelers expect transit routing apps running these algorithms to have minimal response times. Thus, the main goal of our work is to extend PTR algorithms and make them more efficient and practical.

\subsection{Recent approaches}

Popular algorithms for PTR developed in the past decade include---Transfer Patterns \citep{bast2010fast}, Connection Scan Algorithm (CSA) \citep{dibbelt2013intriguingly}, Round-based Public Transit Routing (RAPTOR) \citep{delling2015round}, and Trip-Based public Transit Routing (TBTR) \citep{witt2015trip}. RAPTOR computes Pareto-optimal journeys by increasing the number of allowable transfers in each round. It works directly on a timetable and does not require any preprocessing. It has been extended to include multiple criteria (e.g., arrival time, transfer, fare zones, and reliability). CSA also works directly on a timetable but employs a lightweight preprocessing scheme to store all the connections in a sorted array. Its performance is comparable to RAPTOR for small to medium-sized networks. However, as the network size increases, CSA's performance is poorer because a large number of connections must be evaluated before the algorithm terminates. Transfer Patterns is another efficient bicriterion transit routing algorithm built on the idea that the optimal journeys for a source-destination pair can be derived from a small fixed subset of nodes. Given this set, a significant portion of the network can be ignored during the query stage. However, the preprocessing associated with the Transfer Patterns algorithm is much higher than its counterparts. TBTR, on the other hand, involves a lightweight preprocessing procedure and shows slightly better performance than RAPTOR and CSA. Another line of research in PTR involves modeling online routing strategies \citep{hall1986fastest, hickman1997transit, nguyen1998implicit, khani2015trip, chen2015optimal, li2015finding, rambha2016adaptive, khani2019online}. These are particularly useful in situations involving transfers and uncertain trip times. In this paper, we focus on deterministic settings.

\begin{figure}[htbp]
 \centerline{\includegraphics[scale=0.38]{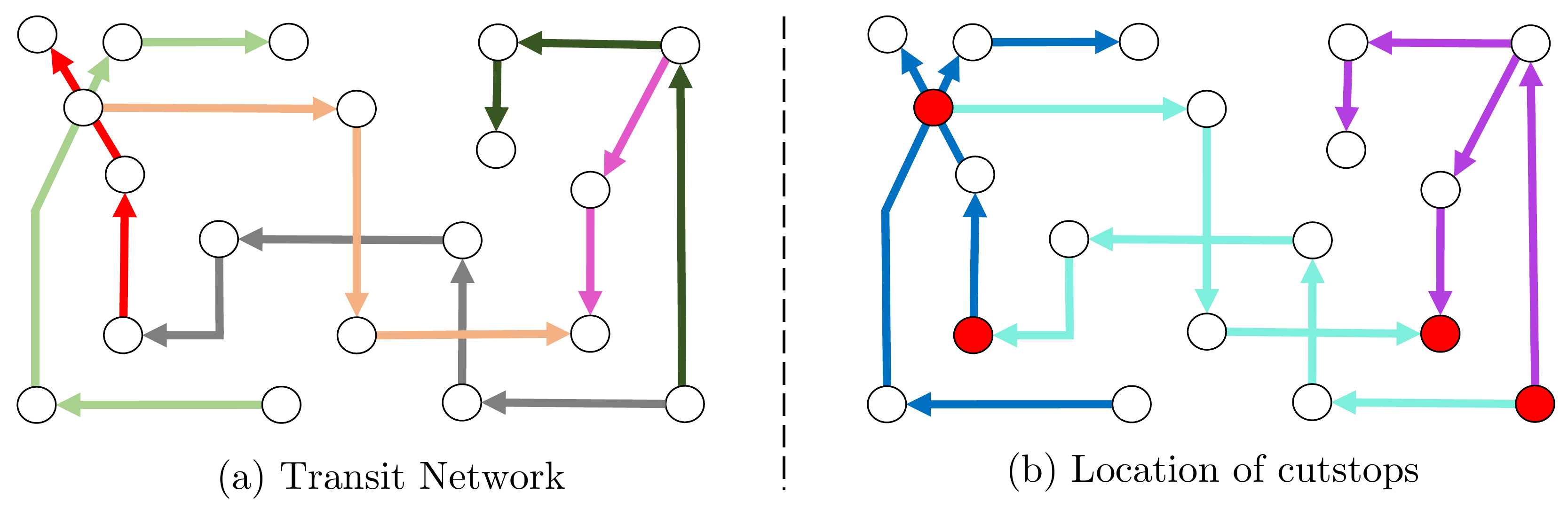}}
     \caption{Partitioning techniques used in PTR algorithms}
     \label{fig:idea}
 \end{figure}

A more recent field of study in PTR is to use partitioning-based approaches to improve query times. Transit networks are partitioned (either based on routes or stops) into subnetworks and the optimal paths between the boundary nodes are precomputed. For example, consider the transit network in the left panel of Figure \ref{fig:idea} where each route is indicated by a different color. The panel on the right shows the network split into three subnetworks. For source-destination pairs that lie completely within each subnetwork, we can use regular PTR algorithms on a smaller graph. However, to travel between these subnetworks, a passenger has to pass through the red boundary stops (also known as cutstops). Thus, we can accelerate routing queries by precomputing and storing the shortest paths between the red stops for all departure times. Building on this idea, CSA was extended to ACSA in \cite{strasser2014connection}, Transfer Patterns was extended to Scalable Transfer Patterns in \cite{bast2016scalable}, and RAPTOR was extended to HypRAPTOR in \cite{delling2017faster}. To the best of our knowledge, ACSA is the only algorithm (relevant to the current study) to exploit a multilevel paradigm. However, its application is limited since it only solves the Earliest Arrival Problem. For a more detailed discussion on CSA and related algorithms, refer \cite{strasser2017algorithm} and \cite{grotschla2017complexity}. Although the preprocessing times of Scalable Transfer Patterns is less than that of Transfer Patterns, it is still higher when compared with other PTR approaches. A partitioning variant for TBTR has not been explored till now.

As stated above, partitioning-based speed-up methods find and store the optimal paths between a given set of cutstops. To do this, existing approaches repeatedly apply a One-To-One algorithm (e.g., rRAPTOR, rTBTR) for all possible cutstop combinations. This step is the major bottleneck during preprocessing. \cite{sauer2020efficient} solved this issue by proposing a One-To-Many framework by combining UnLimited TRAnsfer technique (ULTRA) with Parallel Hardware-Accelerated Shortest path Trees (PHAST). However, adapting the ULTRA-PHAST framework to profile search remains unexplored. Apart from aiding preprocessing, One-To-Many algorithms have several other practical benefits. These include queries involving Points-of-Interest \citep{delling2014customizable}, e.g., \say{find the closest restaurants near me}, transit assignment problems, and building isochrones \citep{bauer2008computing, baum2016fast}, i.e., a set of vertices reachable from a given point within a time or distance limit. While several extensions to road routing algorithms have been proposed to handle such problems \citep{knopp2007computing, delling2013phast}, literature on One-To-Many PTR algorithms is sparse to date.

\subsection{Research gaps and contributions}
Table \ref{tab:research_gaps} summarizes the gaps in existing frameworks for transit routing and highlights the features of the algorithms proposed in this paper. 

\begin{table}[h]
\centering\renewcommand\cellalign{c}
\setcellgapes{3pt}\makegapedcells
    \caption{Research gaps in PTR literature}
    \begin{center}
\begin{adjustbox}{width=1\textwidth}
    \begin{tabular}{c c c c c}
    \hline
    \makecell{\textbf{{\large Original} }\\ \textbf{{\large framework}}} &  
    \makecell{\textbf{{\large One-To-Many}} \\\textbf{{\large version}}} & \makecell{\textbf{{\large Standard}}\\\textbf{{\large partitioning}}} & \makecell{\textbf{{\large Multilevel}}\\\textbf{{\large partitioning}}} & \makecell{\textbf{{\large Other}}\\\textbf{{\large extensions}}}\\
    \hline
    \makecell{{\large Transfer}\\{\large Patterns}\\\cite{bast2010fast}}& $-$  & \cite{bast2016scalable} & $-$ & \cite{bast2014frequency}\\
    \hline
    \makecell{{\large CSA}\\\cite{dibbelt2013intriguingly}}   & $-$ & \cite{strasser2014connection} & \cite{strasser2014connection} &$-$\\
    \hline
    \makecell{{\large RAPTOR}\\\cite{delling2015round} } &  Appendix \ref{sec:raptor_appen} & \cite{delling2017faster} & {\large Section \ref{subsec:MHypTBTR}} & \cite{delling2019fast}\\
    \hline
    \makecell{{\large TBTR}\\\cite{witt2015trip} } &   {\large Section}  \ref{sec:OneToMany}  &  {\large Section} \ref{sec:reduce_querytime}   &  {\large Section \ref{subsec:MHypTBTR}}  & \makecell{ \cite{witt2016trip} \\ \cite{McTBTR}\\ \cite{sauer_et_al}\\ \cite{lehoux_et_al:OASIcs:2020:13139}\\
    }\\
    \hline
\end{tabular}
\end{adjustbox}
\label{tab:research_gaps}
\end{center}
\end{table}

More specifically, the following contributions address these gaps. The improvements summarized here are based on empirical results from six country- and city-level transit datasets. Results from Switzerland, Netherlands, and Sweden were found to be very encouraging are summarized below.
\begin{itemize}
    \item HypTBTR: Section \ref{sec:reduce_querytime} proposes HypTBTR by combining TBTR with a partitioning-based speed-up method. For One-To-One shortest path queries, HypTBTR was found to be 23--37\% faster than the TBTR algorithm. Additional analysis studying the effect of different weighting schemes and hypergraph partitioning tools is also presented. 

    \item One-To-Many rTBTR: Solving profile queries is one of the major bottlenecks in modern PTR approaches such as Scalable Transfer Patterns, HypRAPTOR, and HypTBTR. To address this problem, we extend the popular rTBTR algorithm to its One-To-Many variant in Section \ref{sec:OneToMany}. This algorithm not only reduces the preprocessing times significantly, but can also help query the shortest path between two locations (which may have multiple bus stops near them) instead of two transit stops. Compared to existing approaches, our implementation was found to be 90--95\% faster.  

    \item Multilevel extensions for HypTBTR and HypRAPTOR: The reduced query times in HypTBTR and HypRAPTOR come at the cost of increased preprocessing. We solve this issue in Section \ref{subsec:MHypTBTR} using multilevel partitioning. Compared to standard partitioning implementations, preprocessing times reduced by approximately 5--53\%.

\end{itemize}

However, these results are not generalizable. We also call attention to other transit networks---Israel, Taichung, and Bangalore---that do not benefit from partitioning in Section \ref{sec:Experiments} and point to a few network topology-specific metrics that can distinguish these instances from the success stories. 
    
The remainder of the paper is structured as follows. In Section \ref{sec:pre}, we introduce terminology related to PTR, and review TBTR and RAPTOR. In Section \ref{sec:reduce_querytime}, we reduce the TBTR algorithm's query time by extending it to HypTBTR. Section \ref{sec:preprocessing} is devoted to reducing preprocessing times. Specifically, Subsection \ref{sec:OneToMany} introduces the One-To-Many rTBTR algorithm. Subsection \ref{subsec:MHypTBTR} showcases how multilevel partitioning can be used to further reduce the preprocessing in HypTBTR and HypRAPTOR. In Section \ref{sec:Experiments}, we compare the performance of the proposed techniques on a few transit datasets. Finally, Section \ref{sec:Conclusions} summarizes our findings and suggests potential extensions to the current research. 

To keep the paper concise, other algorithmic details and results from additional experiments are provided in the appendices. Appendix \ref{sec:tbtr_preprocessing} contains the steps involved in TBTR preprocessing. In Appendix \ref{sec:raptor_appen}, we present the One-To-Many variant of rRAPTOR. Appendix \ref{sec:weighting_scheme_compari} illustrates the effect of different hypergraph weighting schemes. Lastly, Appendix \ref{sec:hmetis_vs_kahypar} compares the performance of different partitioning methods---hMETIS, KaHyPar, and an Integer Program (IP). 

\section{Preliminaries}
\label{sec:pre}

We describe the terminology used in the public transit networks in Subsection \ref{subsec:Terminologies} and summarize the notation in Table \ref{tab:glosssary}. Subsection \ref{subsec:RAPTOR_TBTR} reviews a few journey planning algorithms that are closest to our work.

\subsection{Terminology}
\label{subsec:Terminologies}
A \textit{timetable} in a public transit network is defined as a tuple $(S, R, T, F)$, where $S$ is a set of \textit{stops}, $R$ denotes the set of \textit{routes}, $T$ represents the set of \textit{trips}, and $F$ indicates a set of \textit{footpaths}. The timetable corresponds to bus/train schedules and contains information related to a transit network and the movement of vehicles on routes. The most common format for representing timetable information is \textit{General Transit Feed Specification} \href{https://developers.google.com/transit}{(GTFS)}. Maintained by Google, it defines headers for multiple files and rules on how they are related to each other.

A \textit{stop} $s\in S$ is a distinct location in the network where the passengers can board/alight vehicles. For example, bus or train platforms. The source and destination stops are labeled $\source$ and $\target$, respectively. A \textit{route} $r\in R$ is a sequence of stops followed in a particular order. The $i^{th}$ stop on route $r$ is denoted by $\routestop$. The length of route $r$, denoted by $\routelen{r}$, is the total number of stops in the route.

A \textit{trip} $t \in T$ is defined as the movement of a vehicle along a route. A \textit{stop event} (or simply an \textit{event}) refers to a trip arrival or departure at a stop. Similar to $\routestop$ and $\routelen{r}$, we use $\tripstop{t}{i}$ and $\triplen{t}$ to denote $i^{th}$ stop and the length/number of stops of trip $t$, respectively. A trip segment $\tripsegment{t}{i}{j}$ is used to refer to a section of a trip $t$ from stop index $i$ to $j$, that is from the $i^{th}$ stop of the trip to the $j^{th}$ stop. Route ID of trip $t$ is indicated by $\triproute{t}$. The arrival and departure times at the $i^{th}$ stop of trip $t$ are denoted by $\triparr{t}{i}$ and $\tripdep{t}{i}$, respectively. Unless stated otherwise, trips along a route are assumed to be non-overtaking, i.e., trips follow the First-In-First-Out (FIFO) property. Additionally, for two trips $t_1$ and $t_2$ operating on the same route, i.e., $\triproute{t_1}=\triproute{t_2}$, the following notation is used to indicate precedence. If $\triparr{t_1}{i}< \triparr{t_2}{i}\, \forall~i = 1,\ldots, \triplen{t_1}$, $t_1\prec t_2 $. Likewise, if $\triparr{t_1}{i}\leq \triparr{t_2}{i}\, \forall~i = 1,\ldots, \triplen{t_1}$, we express this scenario as $t_1\preceq t_2$.

A footpath $\footstops{s_1}{s_2} \in F$ (also sometimes referred to as a \textit{transfer}) is a pedestrian connection between stops $s_1, s_2 \in S$. For a footpath $\footstops{s_1}{s_2} \in F$, the time required to travel between stops $s_1$ and $s_2$ is denoted by $\footpath{s_1}{s_1}$. We assume that footpaths are transitively closed, i.e., if $\footstops{s_1}{s_2} , \footstops{s_2}{s_3} \in F$, then $\footstops{s_1}{s_3} \in F$. Also, footpath times are assumed to follow the triangle inequality, i.e.,  $\footpath{s_1}{s_3}\leq\footpath{s_1}{s_2}+\footpath{s_2}{s_3}$. Since in GTFS, the \textit{transfers.txt} file that contains footpath details between stops is optional, most online sources do not provide it. Thus, to construct the set $F$, we first define an upper limit on the walking time between a pair of stops. Additional footpaths are then added so that $F$ is transitively closed. We also define the neighborhood of a stop $s$, denoted by $\neighbourhood{s}$, as $\neighbourhood{s}=\{s'|\hspace{5pt}\footstops{s}{s'}\in F\}\cup\{s\}$. Intuitively, for a stop $s$, $\neighbourhood{s}$ is a set of stops directly connected to $s$ via footpaths in $F$ and includes itself. 

For processing shortest path queries efficiently, we need to precompute trip-transfers. A \textit{trip-transfer}, denoted by $\triptrans{t_1}{i}{t_2}{j}$, represents a possible transfer from $i^{th}$ stop of trip $t_1$ to $j^{th}$ stop of trip $t_2$. Note that transfer and trip-transfer are different terms and are not interchangeable. While a transfer just refers to a footpath connection between two stops, a trip-transfer indicates switching between two trips. The switch is possible by either alighting the current trip and waiting at the same stop to board another trip or by walking to a different stop via a footpath and then boarding a trip. The set of trip-transfers is denoted by $\Tset$. See Section \ref{subsec:RAPTOR_TBTR} and Appendix \ref{sec:tbtr_preprocessing} for more details on computing $\Tset$.

A \textit{journey} $y$ is a sequence of trips and footpaths (in the order of traversal). To evaluate a journey $y$, we define
a vector $g(y)=(g_{1}(y),g_{2}(y),\ldots,g_{m}(y))$, where each element of the right-hand side represents the attributes of various optimization criteria associated with $y$. A journey $y_1$ dominates journey $y_2$ if all the elements $g(y_1)$ are no worse than that in $g(y_2)$. For example, if the arrival time and the number of transfers are the optimization criteria, a journey $y_1$ with $g(y_1)=$ (1030, 1 transfer)  dominates a journey $y_2$ with $g(y_2)=$ (1100, 2 transfers). The set of $g(y)$ for different Pareto-optimal journeys is denoted by $\Jset$. A set $\Jset$ is \textit{Pareto-optimal} if none of the journeys in $\Jset$ are dominated by the others. For example, $\Jset=\big\{(1030$, $1$ transfer), (1100, $0$ transfers$)\big\}$ is Pareto-optimal. 

Some studies also consider the \textit{minimum change time} and \textit{dwell time} at each stop. Minimum change time refers to the time spent by a passenger while transferring between trips. Dwell time represents the time interval for which bus/train waits at a stop before departing. To keep the pseudocodes simple, both these parameters are assumed to be zero for all stops, but extending them to a realistic setting is straightforward.

\begin{longtable}[h]{p{2cm} p{13cm}}
    \caption{Glossary\label{tab:glosssary}}\\
        \hline
    \textbf{Symbol} & \textbf{Description}\\
        \hline
    \endfirsthead
    \caption{Glossary (continued)}\\
    \hline
    \textbf{Symbol} & \textbf{Description}\\
    \midrule
    \endhead
    $s$ & stop\\
    $S$ & set of all stops\\
    $S_i$ & set of stops belonging to the $i^{th}$ partition\\
    $\source,\target$ & source, destination stop\\
    &\\
    $\footstops{s_1}{s_2}$ & footpath connection between stop $s_1$ and $s_2$\\
    $F$ & set of all footpath connections\\
    $\footpath{s_1}{s_2}$ & duration of $\footstops{s_1}{s_2}$\\
    &\\
    $t$ & trip\\
    $T$ & set of all trips\\
    $\tripstop{t}{i}$ & $i^{th}$ stop of trip $t$\\
    $\tripflag{t}$ & trip flag of trip $t$\\
    $\tripindex{t}$ & index of the first stop of trip $t$ that has been scanned in TBTR's query phase\\
    $\triproute{t}$ & route of trip $t$\\
    $\triparr{t}{i}$  & arrival time of trip $t$ at stop index $i$\\
    $\tripdep{t}{i}$  & departure time of trip $t$ at stop index $i$\\
    $\tripsegment{t}{i}{j}$    & trip-segment of trip $t$ from stop index $i$ to $j$ \\           
    $\triptrans{t_1}{i}{t_2}{j}$    & possible transfer between $i^{th}$ stop of trip $t_1$ and $j^{th}$ stop of trip $t_2$\\
    $\triplen{t}$ & total number of stops in trip $t$ \\
    &\\
    $r$ & route\\
    $R$ & set of all routes\\
    $R_i$ & set of routes belonging to $i^{th}$ partition\\
    $\routestop$   & $i^{th}$ stop on route $r$\\
    $\routelen{r}$ & total number of stops in route $r$\\
    &\\
    $\dep$ & departure time\\
    $\maxtrans$ & maximum allowed transfers\\
    $p$ & number of partitions\\
    &\\
    $\Lset$   & set of tuples of form $(\route, i, \Delta \tau)$ s.t. $\target$  can be reached from $\routestop$ in $\Delta \tau$ time\\
    $\Tset$     & trip-transfers set\\  
    $\Jset$     & Pareto set of tuples $(\tau_{opt}(n), n)$ where $\tau_{opt}(n)$ is the earliest arrival time at the destination using at most $n$ transfers \\
    $\fillintrip$   & set of trips required for fill-in\\
    $\neighbourhood{s}$   & neighborhood of $s$\\
    \hline
    
\end{longtable}

\subsection{RAPTOR and TBTR} 
\label{subsec:RAPTOR_TBTR}
\begin{figure}[htbp]
 \centerline{\includegraphics[scale=0.32]{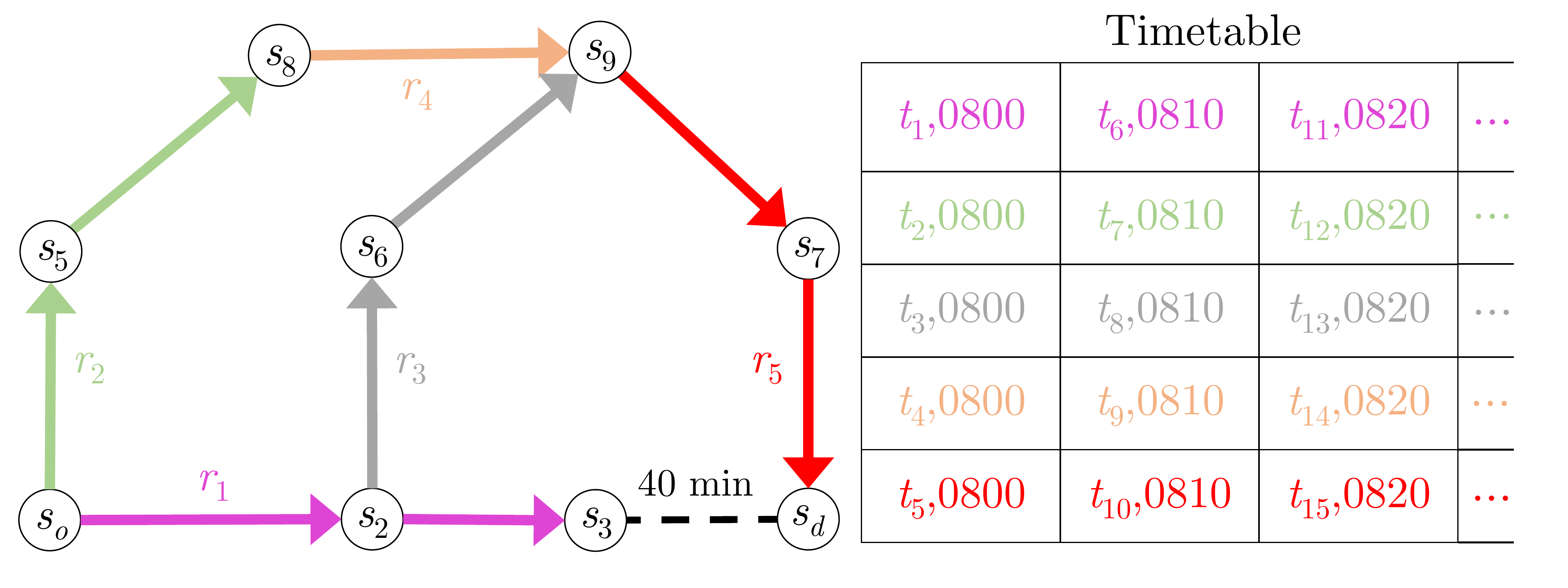}}
     \caption{Toy example to illustrate PTR algorithms}
     \label{fig:test_net}
 \end{figure}

For illustrating the algorithms discussed in this paper, we use the toy network shown in Figure \ref{fig:test_net}. The colored solid edges indicate different routes. The timetable is periodic, i.e., there is a trip on each route starting from 0800 at a 10-minute frequency. Each row represents a route (color-coded) and every cell is a tuple (trip ID, starting time). Let the travel time between any two stops directly connected by an edge be 10 minutes. The dashed line $s_3$--$\target$ represents a footpath with a travel time of 40 minutes. Let the source, destination, and departure time be $\source$, $\target$, and 0800, respectively.

\subsubsection{RAPTOR} 
RAPTOR \citep{delling2015round} is a fully dynamic routing algorithm mainly designed for PTR. While the basic version of the algorithm minimizes the bicriterion problem involving travel time and the number of transfers, its extensions can incorporate other optimizing criteria such as the number of fare zones and cost. RAPTOR uses array and bag data structures to store the transit timetable in the form of trips, and works in rounds. The $n^{th}$ round computes arrival times at stops reachable using exactly $n-1$ transfers. For a stop $s$, two types of labels are used: $\tau_{opt}(n,s)$ which denotes the earliest arrival time label in round $n$ i.e., using exactly $n-1$ transfers and $\tau^*(s)$ which represents the best arrival time at $s$ so far. (Note that unlike RAPTOR, the TBTR algorithm and the rest of the paper assumes that $\tau_{opt}(n,s)$ represents the optimal journey times using \textit{at most} $n$ transfers.) Each round is divided into two phases: a \textit{route phase} and a \textit{transfer phase}. 

In the \textit{route phase}, all routes serving a set of marked stops (source stop or stops whose label was improved in the previous round) are collected. For each of these routes, the earliest possible trip that can be boarded from its marked stop is determined, and its subsequent stop labels are improved (if possible). Any stop whose label is updated is added to the marked stop list. In the \textit{transfer phase}, footpath connections of stops in the marked stop list are evaluated. Stops whose arrival time improves with these footpath connections are added to the marked stop list. RAPTOR has also been extended to handle range queries (rRAPTOR) and multi-criteria queries (McRAPTOR). The following example illustrates the RAPTOR algorithm. The pseudocode for RAPTOR is reviewed in Appendix \ref{sec:raptor_appen}.

In Figure \ref{fig:test_net}, in Round 0, the label $\tau_{opt}(0,\source)$ is updated to the departure time $\dep$ and is added to the marked stop list. Thus, the first round involves boarding the earliest possible trips on the pink $(r_1)$ and green $(r_2)$ routes at 0800 and updating its subsequent stop labels. Stops whose labels are updated (i.e., $s_{2}, s_{3}, s_{5},$ and $s_{8}$), are added to the marked stop list. In the transfer phase of Round 1, footpath connections are evaluated, and the label of stop $\target$ is updated to 0900, and $\target$ is added to the marked stop list. Next, Round 2 begins its route phase using the marked stop list from the transfer phase of Round 1. The algorithm proceeds similarly and terminates when the maximum transfer limit is reached or no stop labels are updated. For the current example, Round 1 explores the pink route $(r_1)$, green route $(r_2)$, and the footpath; Round 2 scans the grey $(r_3)$ and orange route $(r_4)$; and finally, Round 3 explores the red route $(r_5)$ and terminates. The final Pareto-optimal journeys are (0900, 0 transfers) and (0850, 2 transfers).

\subsubsection{TBTR}
\label{sec:TBR_main}
TBTR \citep{witt2015trip} solves the bicriterion (travel time and transfer) optimization problem using trips and trip-transfers as building blocks. The core idea in TBTR is similar to a breadth-first search on a graph in which trips are modeled as nodes and trip-transfers as edges. The first level can be imagined to represent the earliest trips that can be boarded from the source stop, and each subsequent level corresponds to a transfer. TBTR's search pattern is similar to RAPTOR, but unlike RAPTOR, it does not maintain multi-labels for all stops. The \textit{Preprocessing phase} and \textit{Query phase} of TBTR are as described below.

\textit{Preprocessing phase}: TBTR has a two-stage preprocessing phase. Using the GTFS set, the first stage collects all possible trip-transfers in a set $\Tset$ (referred to as the \textit{Trip-transfers set}). This may result in a large collection of trip-transfers, most of which are never part of optimal journeys for any source-destination pair. Hence, the second stage aims to reduce $\Tset$ to only contain useful trip-transfers. Appendix \ref{sec:tbtr_preprocessing} provides the pseudocodes for both the stages along with relevant illustrations. 

\textit{Query phase}: Given an input ($\Tset$, $\source$, $\target$, $\dep$), the query stage maintains a set $\Jset$ of Pareto-optimal tuples of the form (arrival time, number of transfers) and a queue $Q_n$ of trips-segments to be scanned for each allowed number of transfers $n$, where $n\in \{0,1,\ldots, \maxtrans\}$ and $\maxtrans$ is the maximum transfer limit.  $Q_0$ is initialized with the earliest trip-segments that can be boarded on all routes passing through stops in $\neighbourhood{\source}$. Additionally, for every trip $t$, we maintain a variable, $\tripindex{t}$, which denotes the index of the first stop of trip $t$ that can be reached. To efficiently check if a trip passes through $\target$, the algorithm initializes a set $\Lset$ which keeps track of all routes that pass through at least one of the stops in $\neighbourhood{\target}$. The best known arrival time at the destination stop $\target$ using at most $n$ transfers is denoted by $\tau_{opt}(n)$. TBTR, like RAPTOR, also works in rounds. In the $n^{th}$ round, trip-segments in the queue $Q_n$ are scanned. While scanning a trip-segment $\tripsegment{t}{h}{k}$ in round $n$, the following operations are performed:

\begin{itemize}
    \item Using $\Lset$, check if the trip $t$ passes through stop $\target$ (or stops in its neighborhood) and results in a non-dominated label. If so, update $\tau_{opt}(m)~\forall~ m=n,\ldots,\maxtrans$. Note that reducing the labels for all subsequent rounds ensures that the optimality check in later rounds is done against the destination's best arrival time. 

    \item Using the trip-transfers $\triptrans{t}{i}{t^\prime}{j}$ available from the $i^{th}$ stop of $t$ within a trip-segment $\tripsegment{t}{h}{k}$, we add trip-segments $\tripsegment{t^\prime}{j}{\tripindex{t^\prime}}$ to the queue $Q_{n+1}$ if the arrival time at the $(h+1)^{th}$ stop  on $t$ is less than the destination's best known arrival time (i.e., $\triparr{t}{h+1}<\tau_{opt}(n)$) and the trip $t^\prime$ from $j$ onwards has not already been explored (i.e., $j<\tripindex{t^\prime}$). 
\end{itemize}

\begin{figure}[h]
\centerline{\includegraphics[scale=.9]{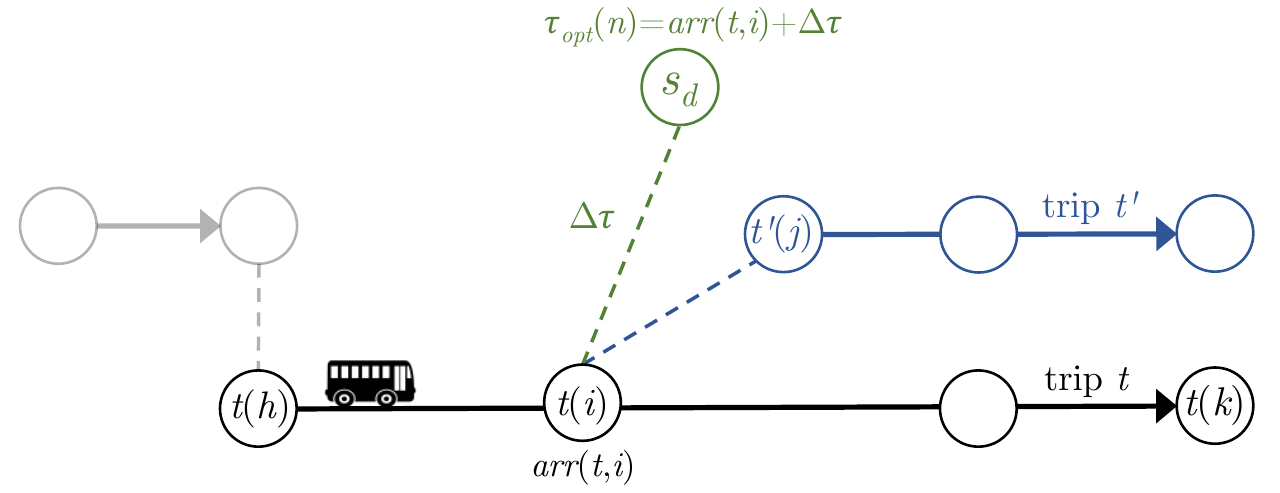}}
\caption{Illustration of the \textit{while}-loop in the TBTR algorithm. The green and blue portions reflect operations carried out in Lines 16--19 and 21--23 of Algorithm \ref{alg:TBTR}, respectively.}
\label{fig:whileabj}
\end{figure}

The algorithm stops when the maximum transfer limit $\maxtrans$ is reached, i.e., $n=\maxtrans$ or when $Q_n$ is empty for $n < \maxtrans$. The pseudocode for the TBTR's bicriterion query is shown in Algorithm \ref{alg:TBTR}. Lines 1--5 represent the initialization phase where $\tau_{opt}(n)$ is initialized to $\infty$ for all $n\leq\maxtrans$. Lines 6--8 generate the set $\Lset$, which is used during the query phase to check if a journey reached the destination stop. Lines 9--14 add the trip-segments that can be boarded from stops in $\neighbourhood{\source}$ to $Q_0$. Lines 15--26 scan every trip-segment $\tripsegment{t}{h}{k}$ as described above. Connections from the trip-segment (assuming the condition in Line 21 is satisfied) are collected in a list \textit{clist}. Next, Line 24 calls the function \textsc{Enqueue} which iterates over the elements of \textit{clist} and adds unexplored trip-segments to $Q_{n+1}$. Figure \ref{fig:whileabj} illustrates the main \textit{while}-loop of the TBTR algorithm. 

Consider the example from Figure \ref{fig:test_net} to illustrate the TBTR algorithm and assume $\Tset=$ $\big\{\triptrans{t_1}{2}{t_8}{1},$ $\triptrans{t_2}{3}{t_{14}}{1},$ $\triptrans{t_{14}}{2}{t_{20}}{1},$ $\triptrans{t_8}{3}{t_{20}}{1}\big\}$. Note that this is smallest $\Tset$ required to solve the example query. Round 0 starts by scanning the trips in queue $Q_0 = \{\tripsegment{t_1}{1}{3},\tripsegment{t_2}{1}{3}\}$. Since trip $t_1$ is connected to $\target$ via a footpath, $\tau_{opt}(0)$ is updated to 0900. Next, we scan $\Tset$ to get trip-transfers from trips $t_1$ and $t_2$ and add the corresponding trip-segments to queue $Q_2$. Round 1 starts with $Q_1=\{\tripsegment{t_8}{1}{3}, \tripsegment{t_{14}}{1}{2}\}$ and proceeds in a similar fashion. For Round 2, $Q_2=\{\tripsegment{t_{20}}{1}{3}\}$. The final output of the algorithm is $\tau_{opt}(0)=0900$, $\tau_{opt}(1)=0900$, and $\tau_{opt}(2)=0850$. 

TBTR has been extended to incorporate range queries, similar to rRAPTOR. The idea behind this algorithm is to run the main query for different departure times while preserving labels between runs. Later departure times are processed first. Another version of TBTR is Trip-based Routing using Condensed Search Trees or TBTR-CT \citep{witt2016trip}. This algorithm exploits the observation that the optimal journey for a source-destination pair at any time can be constructed using only a fixed set of routes instead of the complete network. It preprocesses routes for each source-destination pair, and thus, the query phase explores a much smaller graph, resulting in faster queries. Just like Transfer Patterns \citep{bast2010fast}, faster query times in TBTR-CT come at the cost of high preprocessing time and increased memory usage. 
\cite{McTBTR} proposed two new variants of TBTR: the Walking TBTR, which additionally minimizes the walking time along a journey, and Fare Zone TBTR, which optimizes the fare zones as a third criterion. Their results have been compared with McRAPTOR. \cite{sauer_et_al} extended TBTR to handle multi-modal bicriterion routing problems by combining it with ULTRA \citep{baum2019unlimited}.

\begin{algorithm}[t]
\caption{TBTR: Bicriterion query}
\label{alg:TBTR}
 \hspace*{\algorithmicindent} 
    \textbf{Input}: GTFS, $\Tset$, $\source$,  $\target$, $\dep$, $\maxtrans$\\
 \hspace*{\algorithmicindent} 
    \textbf{Output}: $\Jset$
    \begin{algorithmic}[1]
	\vspace{2mm}
	    \State $n\leftarrow 0$
		\State $\Jset, \Lset \leftarrow \emptyset$
		\State $\tau_{opt}(n)\leftarrow  \infty \forall \, n=0, 1, \ldots,\maxtrans$ 
	    \State $Q_n \leftarrow \emptyset \, \forall \, n=0, 1, \ldots,\maxtrans$
        \State $\tripindex{t} \leftarrow \triplen{t} \hspace{3pt}\forall$ $t \in T$ 

    \vspace{2mm}
        \For {$s \in \neighbourhood{\target}$}
            \For { $(r,i)$ s.t. $\routestop=s$}
                \State $\Lset\leftarrow \Lset \cup \{(r,i,\footpath{s}{\target})\}$
        \EndFor
        \EndFor
    \vspace{2mm}
        \State $clist \leftarrow \emptyset$
        \For {$s \in \neighbourhood{\source}$}
            \For { $(r,i)$ s.t. $\routestop=s$}
                \State $t\leftarrow$ earliest trip on $r$ s.t. $\dep + \footpath{\source}{s} \leq \tripdep{t}{i}$
                \State $clist\leftarrow clist \cup \big\{(t,i,n)\big\}$
        \EndFor
        \EndFor
        \State\textsc{Enqueue}$(clist, \textit{ind}, Q_n)$ 
    \vspace{2mm}

        \While{$Q_n \neq \emptyset$ \AND $n\leq\maxtrans$}
            \For {$\tripsegment{t}{h}{k} \in Q_n$}
                    \For {$(\triproute{t}, i , \Delta\tau)\in\Lset$ s.t. $h<i\leq k$ \AND $\triparr{t}{i} + \Delta\tau<\tau_{opt}(n)$}
                    \For{$m =n, n+1,\ldots,\maxtrans$}
                        \State $\tau_{opt}(m)\leftarrow\triparr{t}{i} + \Delta\tau$
                    \EndFor
                    \EndFor
                    \State $clist \leftarrow \emptyset$
                    \If{$\triparr{t}{h+1}<\tau_{opt}(n)$}
                    \For {$\triptrans{t}{i}{t^\prime}{j} \in \Tset$ s.t. $h<i\leq k$}
                        \State add $(t^\prime,j,n+1)$ to $clist$ if not already present
                    \EndFor
                    \EndIf
        \State\textsc{Enqueue}$(clist, \textit{ind}, Q_{n+1})$ 
            \EndFor
        \State add $(\tau_{opt}(n), n)$ to $\Jset$ if it is non-dominated
        \State $n\leftarrow n + 1$
        \EndWhile

        \vspace{2mm}
        \setcounter{ALG@line}{0}
        \Procedure{\textsc{Enqueue}}{list of connections $clist$, vector \textit{ind}, queue $Q$}{
            \For{$(t,i,n)\in~ clist$}
            \If{ $i<\tripindex{t}$}
                \State $Q\leftarrow Q \cup \hspace{3pt} \big\{\tripsegment{t}{i}{\tripindex{t}}\big\}$
                \For { trip $t'$ s.t. $t \preceq t'$ \AND $\triproute{t}=\triproute{t'}$}
                    \State $\tripindex{t'}\leftarrow \min\big\{\tripindex{t'}, i\big\}$
                \EndFor
    		\EndIf
            \EndFor
        
        }
        \EndProcedure

	\end{algorithmic}
\end{algorithm}

\section{Reducing query times}
\label{sec:reduce_querytime}
This section extends TBTR to HypTBTR by combining it with partitioning-based approaches. We start with the relevant background in Subsection \ref{subsec:back}, followed by the development of HypTBTR in Subsection \ref{subsec:HypTBTR}.

\subsection{Background}
\label{subsec:back}
\cite{delling2017faster} proposed HypRAPTOR for faster queries by introducing a preprocessing step which partitions routes into $p$ disjoint sets $R_1, R_2, \ldots, R_p$, also known as \textit{route cells}. The central idea in HypRAPTOR is to construct a hypergraph $\mathcal{G}$ in which nodes represent routes, and hyperedges between subsets of nodes represent \textit{intersecting routes}. Two routes are called intersecting if they have at least one stop in common. Footpaths are also treated as routes with two stops. A partitioning algorithm (based on a min-cut approach) is then used to generate route cells. By definition, route cells are mutually exclusive and exhaustive (i.e., $R_i \cap R_j = \emptyset$ for all $i$ and $j$, and $\bigcup_{i=1}^p R_{i}=R$).

\begin{figure}[htbp]
 \centerline{\includegraphics[scale=0.55]{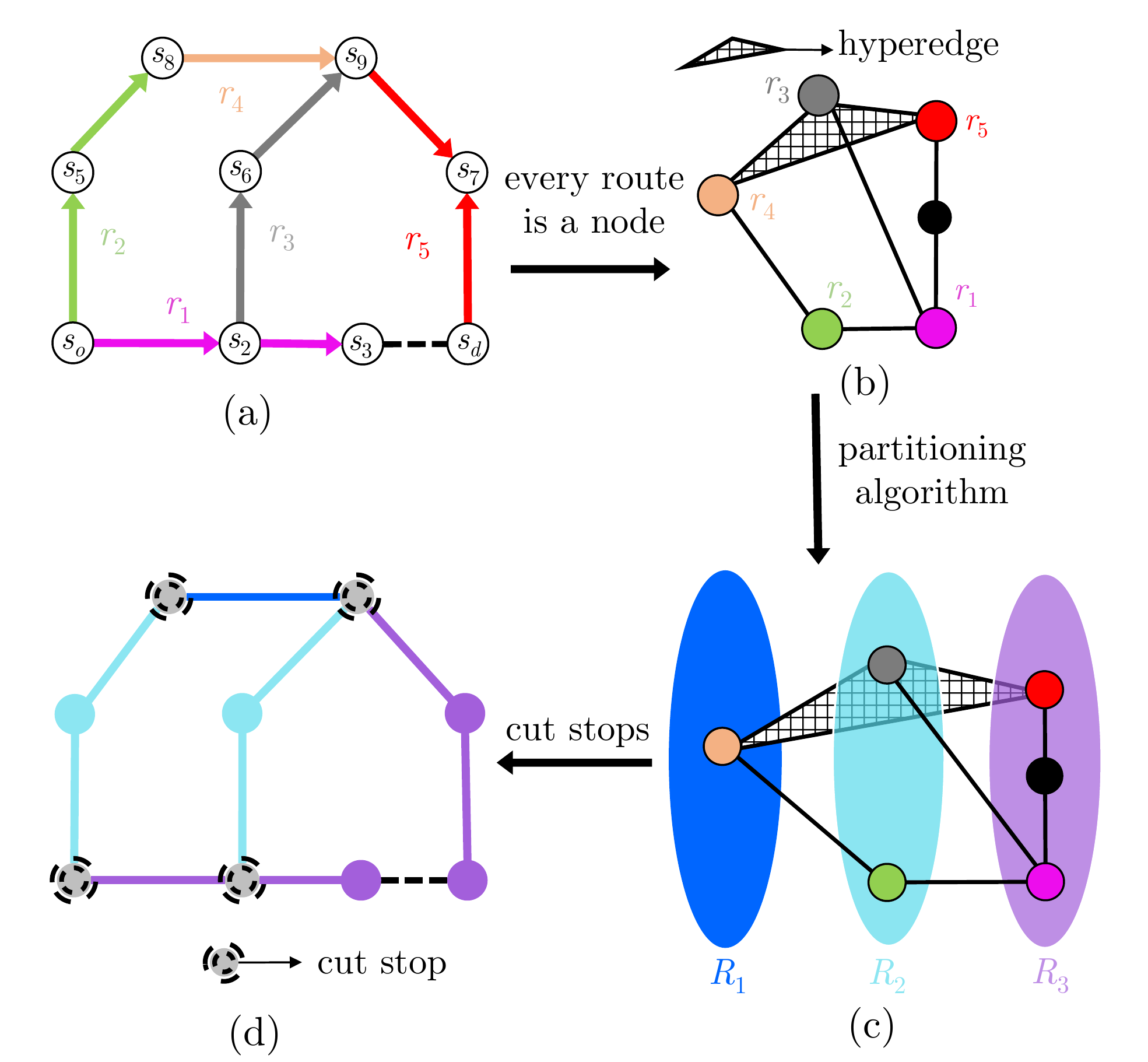}}
     \caption{Route partitioning using a hypergraph representation}
     \label{fig:HYRAPTOR}
 \end{figure}

Suitable edge and node weights can be used to influence the partitions. Multi-edges that result from routes intersecting at more than one stop are reduced to a single edge by summing their weights. Note that every boundary edge of the partition (an edge with nodes in different cells) represents a stop in the original transit network and is referred to as a \textit{cutstop}. Consider the example in Figure \ref{fig:HYRAPTOR}. Subfigure (b) shows the corresponding hypergraph with each route as a node (a hyperedge with three nodes is added between the red, orange, and grey routes). The black node represents the footpath. Assuming $p=3$, a partitioning algorithm creates three route cells namely $R_1$ (dark blue), $R_2$ (cyan), and $R_3$ (purple), as shown in Subfigure (c). Subfigure (d) shows the cutstops derived from the route cells in (c). 

Similar to route cells, \textit{stops cells} $S_0, S_1, \ldots, S_p$ is a partition of stops. To construct a stop cell $S_i$, we start by including all the stops belonging to the routes in the corresponding route cell $R_i$. However, the resulting stop cells are not mutually exclusive due to cutstops. Thus, define $S_0$ to contain all the cutstops and update stop cell $S_i$ as $S_i \leftarrow S_i\hspace{3pt}\backslash\hspace{3pt}S_0$. For example, in Figure \ref{fig:test_net}, the stop cells are $S_0=\{s_0,\ s_2,\ s_8,\ s_9\}$, $S_1=\emptyset$, $S_2=\{s_5,\ s_6\}$, and $S_3=\{s_3,\ s_d$, $s_7\}$. Note that for $p$ partitions, there are $p$ route cells and $p+1$ stop cells.

The next step is to find and store optimal routes between these cutstops, referred to as \textit{fill-in}. To this end, a profile query (using rRAPTOR) is made for all possible pairs of cutstops in $S_0$. If a route belongs to an optimal journey between a pair of cutstops, it is added to the fill-in set. 

In the query phase, given the source and destination, the first step is to identify $\originstopcell$ and $\targetstopcell$, the stop cells to which $\source$ and $\target$ belong to, respectively. Let the corresponding route cells be $\originroutecell$ and $\targetroutecell$. If the source stop is a cutstop, then $\originstopcell$ and $\originroutecell$ are set to $S_0$ and $\emptyset$, respectively. The same convention is used for the destination stop. The algorithm scans a route only if it belongs to the fill-in set or if all its stops are in the source or destination cells. Building on this idea we construct the HypTBTR algorithm. A new hypergraph weighting scheme is proposed, and changes to the query phase are suggested for the trip-based setting. In a subsequent section, we discuss methods to make HypTBTR preprocessing more efficient.

\subsection{HypTBTR}
\label{subsec:HypTBTR}
\subsubsection{Preprocessing phase}
\label{subsec:hyp_preprocessing}
Preprocessing in HypTBTR can be grouped into two phases: \textit{trip-transfer phase} and \textit{fill-in computation phase}. The \textit{trip-transfer phase} generates a trip-transfer set $\Tset$ and is similar to TBTR's preprocessing (see Appendix \ref{sec:tbtr_preprocessing} for details).

The \textit{fill-in computation phase} can be further divided into two steps. The first step is similar to that of HypRAPTOR and generates a hypergraph by modeling each route as a node. We then add hyperedges between routes that have common stops. Next, to generate the partitions and cutstops, we use the state-of-the-art KaHyPar algorithm \citep{Schlag2020_1000105953}. KaHyPar works on a min-cut principle and finds nearly equal-sized cells such that the sum of weights of the hyperedges in the cut is minimized. The size of a cell is defined as the sum of the weights of nodes belonging to it. Thus, different weighting schemes, as described below, can be used to influence the partitioning process. 
\begin{enumerate}
    \item Sc$_1$: This represents a simple unweighted scheme where no weights are imposed on nodes and edges. Thus, the algorithm minimizes the number of hyperedges cut. 

    \item Sc$_2$: This scheme was first proposed by \cite{delling2017faster}. Here, the nodes (i.e., routes in the original transit network) are weighted by the total numbers of events in the route. The weights of nodes corresponding to footpath connections are set to zero. Each hyperedge is assigned a weight equal to the logarithm of the total number of events associated with the corresponding transit stop.  

    \item Sc$_3$: In this case, nodes are weighted similar to Sc$_2$. But for a hyperedge corresponding to a stop $s$, the weight is set to the logarithm of the sum of events associated with all the stops in $\neighbourhood{s}$.
\end{enumerate}

In HypRAPTOR's (or HypTBTR's) preprocessing, walking from the source stop must be allowed. That is, when finding the trips required to cover optimal journeys between a cutstop pair $s_i$ and $s_j$ in the preprocessing phase, Round 0 would have to start with all the earliest trips from $\neighbourhood{s_i}$ (instead of trips from $s_i$). This is because, although the cutstop $s_i$ is a source stop during preprocessing, it can become an intermediate stop between some $\source$ and $\target$ during the query phase. For example, an optimal journey could be of the form---board the first trip from $\source$ and alight at $s_i$, walk from $s_i$ to some stop $s \in \neighbourhood{s_i}$, take a bus from $s$ to $s_j$, and board the next available trip from $s_j$ and reach $\target$. Thus, walking from the source cutstop $s_i$ must be allowed during preprocessing to maintain the algorithm's correctness. Therefore, the fill-in calculations associated with a stop $s$, depend not only on $s$ but on all the stops in $\neighbourhood{s}$. For this reason, we expect Sc$_3$, which sets the hyperedge weights to the logarithm of the sum of events in $\neighbourhood{s}$, to perform better than Sc$_2$. Experiments in Section \ref{sec:Experiments} are carried out using Sc$_3$. Comparison of other weighting schemes is documented in Appendix \ref{sec:weighting_scheme_compari}. 

For a pre-determined partition size $p$, the output of the first step includes $p$ route cells $R_1, R_2, \ldots, R_p$ and $p+1$ stop cells $S_0, S_1, \ldots, S_p$, where $S_0$ is the set of cutstops. The second step in the \textit{fill-in computation phase} is to find the \textit{fill-in trip set} $\fillintrip$, i.e., the set of trips required for all optimal journeys between the cutstops. For this purpose, \cite{delling2017faster} use a profile query for all possible cutstop permutations by repeatedly applying rRAPTOR. To speed-up this process, we propose a new accelerated version of rTBTR, a One-To-Many rTBTR in Section \ref{sec:OneToMany}. Our implementation significantly outperforms the existing approach by rapidly solving profile queries between all possible cutstops pairs. Next, we discuss a few alternate hypergraph partitioning tools that have also been tested in this paper. 

Hypergraph partitioning is a widely studied problem in graph theory because of its numerous applications. As mentioned earlier, the problem involves dividing the vertices of the hypergraph into a given number of partitions such that the number of hyperedges cut is minimized and the size of each partition is bounded. We review an IP formulation of the problem  \citep{kucar2004hypergraph}, followed by a discussion on other algorithmic approaches. 

Let $\mathcal{G}=(V,E)$ represent an undirected hypergraph where $V$ is the set of nodes and $E$ is the set of hyperedges. Denote using $w_v$ and $w_e$, the weight of a node $v$ and a hyperedge $e$, respectively. The decision variables of the problem include $x_{vi}$, which equals 1 if node $v$ belongs to partition $i$, i.e., $v \in R_i$ and is 0 otherwise and $y_{ei}$, which is 1 if hyperedge $e$ lies inside partition $i$, i.e., if $v \in R_i\ \, \forall \, v \in e$. Note that $y_{ei}=1$ implies that hyperedge $e$ is uncut. Let $\alpha_i$ and $\beta_i$ represent the lower and upper bound on the size of the partition $i$. Recall that the size of a partition is defined as the sum of the weights of hypernodes in it. The IP model for hypergraph partitioning can thus be formulated as follows.
\begin{align}
  \max\quad        & \sum_{e \in E}\sum_{i=1}^{p} w_e y_{ei} \label{eq:objective}\\ 
  \text{s.t.\quad} & \sum_{i=1}^{p} x_{vi}=1 & &\forall\, v \in V \label{eq:cover} \\ 
                   & \alpha_i\le \sum_{v \in V}w_v x_{vi}\le \beta_i &   & \forall \,  i= 1,\dots,p   \label{eq:bound} \\ 
                   &    y_{ei}\leq x_{vi}\hspace{5pt}               &   & v \in V, e \in E,  i=1,\ldots,p \label{eq:linkxy} \\ 
                   & x_{vi} \in \{0,1\}                                            &   & v \in V,  i=1,\ldots,p  \label{eq:booleanx} \\ 
                   & y_{ei} \in \{0,1\}                                            &   & e \in E,  i=1,\ldots,p \label{eq:booleany}  
\end{align}

The objective function \eqref{eq:objective} maximizes the sum of the weights of uncut hyperedges. Equation \eqref{eq:cover} ensures that every node belongs to exactly one partition. Constraint \eqref{eq:bound} imposes bounds on the partition sizes. Constraint \eqref{eq:linkxy} stipulates that for a hyperedge $e$, $y_{ei}$ is 1 if and only if all the nodes in $e$ are in partition $i$. Boolean constraints on the decision variables are enforced in \eqref{eq:booleanx} and \eqref{eq:booleany}.

IP models like the one discussed above have both pros and cons. Their biggest advantage is the ability to easily incorporate different weighted objective functions. 
However, while solving for the optimal partitions, enumerative techniques such as branch-and-cut are used, which tend to be slow in practice and do not scale well. 

To overcome these drawbacks, several alternate heuristics (e.g., iterative methods, genetic algorithms, multilevel methods) have been studied in the literature. See \cite{kucar2004hypergraph} for more details. Among these approaches, multilevel partitioning algorithms such as hMETIS \citep{hmetis} and KaHyPar \citep{Schlag2020_1000105953, schlag2021high} are popular and have been found to be superior compared to other approaches such as PaToH \citep{ccatalyurek2011patoh} and Zoltan \citep{devine2006parallel}.

These multilevel methods typically work in three steps: Coarsening, Initial Partitioning, and Refinement. Coarsening reduces the size of the hypergraph by contracting subsets of nodes, i.e., a set of nodes is replaced by a single vertex. It successively generates smaller hypergraphs such that partitions on the smaller hypergraph are not significantly worse than partitions generated directly on the original hypergraph. The next step involves partitioning the coarsened graph. Finally, in the Refinement phase, the graph is uncoarsened by successively projecting it to back to a finer-level. Several refinement techniques such as the FM algorithm \citep{fiduccia1982linear} are used to improve the quality of partitions without violating user-specified constraints. KaHyPar and hMETIS differ in the algorithms used in these three phases. E.g., for coarsening, hMETIS uses the firstchoice algorithm and KahyPar uses community aware coarsening. The readers can refer to \cite{Schlag2020_1000105953} for more details. All experiments in Section \ref{sec:Experiments} were done using KaHyPar. We used publicly available implementations of KaHyPar \href{https://github.com/kahypar/kahypar}{(github.com/kahypar)} and hMETIS \href{http://glaros.dtc.umn.edu/gkhome/metis/hmetis/download}{(hMETIS Code)}. Only the KaHyPar code offered flexibility in configuring various parameters such as the objective function and seed value. For performance metrics of HypTBTR using hMETIS, refer Appendix \ref{sec:hmetis_vs_kahypar}.

\subsubsection{Query phase}
Depending on the fill-in representation, the query phase can be implemented in many ways. The simplest approach is to scan a route only if it belongs to the source or destination cell, or is a part of the fill-in. Other methods include marking every event that is a part of fill-in computations as opposed to marking the whole route or generating a new overlay graph by copying the events of the source and destination cells and fill-in. 

These methods present a trade-off between speed and memory usage. Experiments by \cite{delling2017faster} show that overlay graphs are slightly better, but other methods have comparable performance and the exact benefits vary depending on the network and partition structure. In this paper, we use the simplest form of fill-in representation, i.e., for every trip $t$, we initialize a one-bit variable known as a \textit{trip flag} denoted by $\tripflag{t}$. A trip's flag is true if it belongs to the source or destination cell or is a part of the fill-in. The pseudocode for the query phase is described in Algorithm \ref{alg:HypTBTR_algo}.

Lines 1--5 are used to initialize the variables and are similar to Algorithm \ref{alg:TBTR}. In Line 6, we introduce a function \textit{\textsc{Labeling}} which, given $\source$ and $\target$, identifies $\originstopcell$ and $\targetstopcell$, the stop cells to which source and destination stop belong, respectively. The corresponding route cells are also identified as $\originroutecell$ and $\targetroutecell$. Although we do not necessarily need $\originstopcell$ and $\targetstopcell$, we include it in the pseudocode since it helps establish a proof of correctness. 

Lines 7--9 set the trip flags for every trip by checking if it belongs to the fill-in set or if its route belongs to $\originroutecell$ or $\targetroutecell$. Lines 10--12 define the set $\Lset$, which stores information on all the routes and footpaths leading to the destination stop. Lines 13--18 consider the first trip on different routes that can be boarded from $\neighbourhood{\source}$ and adds them to $Q_0$. Lines 19--30 contain the main iterations of the algorithm. Trips segments are added to a queue and are scanned as before with an extra condition that the $\tripflag{t}$ variable should be True. \\ 


\begin{breakablealgorithm}
\caption{HypTBTR: Bicriterion query}
\label{alg:HypTBTR_algo}
    \noindent\hspace*{\algorithmicindent}\textbf{Input}: GTFS, $\Tset,\ \source,\ \target,\ \dep,\ \fillintrip,\ \maxtrans,\ p$, Stop cells $\{S_0,\ldots, S_p\}$, Route cells $\{R_1,\ldots, R_p\}$\\
    \noindent\hspace*{\algorithmicindent}\textbf{Output}: Pareto set $\Jset$
    \begin{algorithmic}[1]
	\vspace{2mm}
	    \State $n\leftarrow0$
		\State $\tau_{opt}(n)\leftarrow  \infty \forall \, n=0, 1, \ldots,\maxtrans$ 
		\State $\Jset, \Lset \leftarrow \emptyset, \emptyset$
	    \State $Q_n \leftarrow \emptyset \, \forall \, n=0, 1, \ldots, \maxtrans$
        \State $\tripindex{t}, \tripflag{t} \leftarrow \triplen{t}, \text{False} \hspace{4pt}\forall$ $t\in T$ 
        \State $\originstopcell,\ \targetstopcell,\ \originroutecell,\ \targetroutecell,\ \leftarrow$ \textsc{Labeling}$(\source, \target)$
    \vspace{2mm}
    \For {$t\in T$}
        \If{$t \in \fillintrip$ \OR  $\triproute{t} \in \originroutecell\hspace{3pt} \cup \hspace{3pt} \targetroutecell$}
            \State $\tripflag{t}\leftarrow $ True 
    	\EndIf
    \EndFor
    \vspace{2mm}

        \For {$s\in \neighbourhood{\target}$}
            \For { $(r,i)$ s.t. $\routestop=s$}
                \State $\Lset\leftarrow \Lset \cup \big\{(r,i,\footpath{s}{\target})\big\}$
        \EndFor
        \EndFor
    \vspace{2mm}

        \State $clist \leftarrow \emptyset$
        \For {$s\in \neighbourhood{\source}$}
            \For { $(r,i)$ s.t. $\routestop=s$}
                \State $t\leftarrow$ earliest trip on $r$ s.t. $\dep + \footpath{\source}{s} \leq \tripdep{t}{i}$
                    \State $clist\leftarrow clist\cup \big\{(t,i,n)\big\}$
    \EndFor
        \EndFor
            \State \textsc{Enqueue}$(clist, \textit{ind}, Q_n)$
    \vspace{2mm}
        
        \While{$Q_n \neq \emptyset$ \AND $n\leq\maxtrans$}
            \For {$\tripsegment{t}{h}{k} \in Q_n$}
                    \For {$(\triproute{t}, i , \Delta\tau)\in\Lset$ s.t. $h<i\leq k$ \AND $\triparr{t}{i} + \Delta\tau<\tau_{opt}(n)$}
                    \For{$m =n,n+1,\ldots,\maxtrans$}
                    \State $\tau_{opt}(m)\leftarrow\triparr{t}{i} + \Delta\tau$
                    \EndFor
                     \EndFor
                    \State $clist \leftarrow \emptyset$
                     \If{$\triparr{t}{h+1}<\tau_{opt}(n)$}
                    \For {$\triptrans{t}{i}{t^\prime}{j} \in \Tset$ s.t. $h<i\leq k$ }
                        \State add $(t^\prime,j,n+1)$ to $clist$ if not already present
                    \EndFor
                    \EndIf
        \State\textsc{Enqueue}$(clist, \textit{ind}, Q_{n+1})$ 
            \EndFor
        \State add $(\tau_{opt}(n), n)$ to $\Jset$ if it is non-dominated
        \State $n\leftarrow n + 1$
        \EndWhile
        
        \vspace{2mm}
        \setcounter{ALG@line}{0}
        \Procedure{\textsc{Enqueue}}{list of connections $clist$, vector \textit{ind}, queue $Q$}{
        \For{$(t,i,n) \in clist$}
            \If{$\tripflag{t}$ \AND $i<\tripindex{t}$}
                \State $Q\leftarrow Q \cup \hspace{3pt} \big\{\tripsegment{t}{i}{\tripindex{t}}\big\}$
                \For { trip $t'$ s.t. $t \preceq t'$ \AND $\triproute{t}=\triproute{t'}$}
                    \State $\tripindex{t'}\leftarrow \min\big\{\tripindex{t'}, i\big\}$
                \EndFor
    		\EndIf
                \EndFor
        
        }
        \EndProcedure
        \vspace{2mm}
        \setcounter{ALG@line}{0}
        \Procedure{\textsc{Labeling}}{source stop $\source$, destination stop $\target$}{
            \State{$\originstopcell,\ \targetstopcell,\ \originroutecell,\ \targetroutecell,\ \leftarrow \emptyset,\ \emptyset,\ \emptyset,\ \emptyset$}
            \For {$i= 0,1, \ldots, p$}
                \If{$\source \in S_i$}
                \State{$\originstopcell\leftarrow S_i$}
                \State{$\originroutecell\leftarrow R_i$ if $i\neq 0$ else $\emptyset$}
                \EndIf
                \If{$\target \in S_i$}
                \State{$\targetstopcell\leftarrow S_i$}
                \State{$\targetroutecell\leftarrow R_i$ if $i\neq 0$ else $\emptyset$}
                \EndIf
            \EndFor

        }
        \EndProcedure

	\end{algorithmic}
\end{breakablealgorithm}

\textit{Proof of correctness}: Suppose there exists an optimal journey $y$ between $\source$ and $\target$ whose earliest arrival time and number of transfers are not captured by Algorithm \ref{alg:HypTBTR_algo}. The following cases can arise: (i) $\originstopcell=\targetstopcell=\emptyset$, i.e., $\source$ and $\target$ are cutstops, (ii) $\originstopcell=\targetstopcell  \ (\neq\emptyset)$, i.e., $\source$ and $\target$ belong to the same stop cell, or (iii) $\originstopcell\neq\targetstopcell$, i.e., $\source$ and $\target$ belong to different stop cells. 

In Case (i), as both $\source$ and $\target$ are cutstops, HypTBTR will only scan trips in the fill-in set $\fillintrip$ (since Lines 7--9 set $\tripflag{t}=$ True $ \forall~t \in \fillintrip$). Thus, if $y$ is optimal, the fact that $\fillintrip$ contains all the optimal trips between the cutstops is contradicted. In Case (ii), optimal journeys are either fully contained in the route cell or exit the route cell at some cutstop $s_i$ and enter again at another cutstop $s_j$. The former scenario reduces to a simple TBTR without partitioning and in the latter instance, all the trips from the route cell and $\fillintrip$ are scanned in Lines 7--9 and hence it again contradicts the definition of $\fillintrip$. The arguments for Case (iii) are similar. Further, note that Line 29 restricts $\Jset$ to only contain Pareto-optimal labels and hence it does not contain any extra labels corresponding to sub-optimal journeys.

\section{Reducing preprocessing times}
\label{sec:preprocessing}
In this section, we speed-up preprocessing using a One-To-Many rTBTR which reduces the time required for profile queries (Section \ref{sec:OneToMany}) and multilevel partitioning which reduces the calculations required for the fill-in set computation (Section \ref{subsec:MHypTBTR}).

\subsection{One-To-Many rTBTR}
\label{sec:OneToMany}
As discussed in Section \ref{subsec:hyp_preprocessing}, preprocessing involves solving profile queries between all possible cutstop pairs. Existing approaches use a range algorithm (rRAPTOR or rTBTR) for all possible cutstop pair combinations. In this section, we propose a One-To-Many version to make this step more efficient.

For a give source stop $\source$, let \textit{tlist} represent a list of all  departure times of trips (sorted in descending order) from $\source$ (or $\neighbourhood{\source}$ if walking from the source is allowed). Let \textit{dlist} be list of destination stops. Similar to $\Lset$, $\Jset$, and $\tau_{opt}(n)$ in TBTR, we define $\Lset(s)$, $\Jset(s)$, and $\tau_{opt}(n,s)$ for each $s \in$ \textit{dlist} where $\Lset(s)$ keeps track of routes through $\neighbourhood{s}$,  $\Jset(s)$ stores the optimal journey attributes to $s$, and $\tau_{opt}(n,s)$ is the optimal label of stop $s$ using at most $n$ transfers, respectively. Lastly, to preserve labels between runs, we also store the index of the first stop on trip $t$ that can be reached within $n$ transfers, $\tripindexrange{t}$.

In addition to \textit{dlist}, we also maintain a dummy list \textit{dlist}$^\prime$ and a variable \textit{scope} whose purpose is to prune the list of destination stops as the algorithm proceeds. To understand the idea behind pruning the destination list, consider the toy network shown in Figure \ref{fig:test_net}. For simplicity, imagine that only one trip on the pink and green routes departs from $\source$ at 0800 and suppose $dlist=[s_6, s_d]$. Thus, $tlist= \big[\triparr{t_1}{1}, \triparr{t_2}{1}]$. As discussed in Section \ref{sec:TBR_main}, for Round 0, $Q_{0}=\big\{\tripsegment{t_1}{1}{3}, \tripsegment{t_2}{1}{3}\big\}$. Next, Round 1 starts with $Q_{1}=\big\{\tripsegment{t_8}{1}{3}, \tripsegment{t_{14}}{1}{2}\big\}$ and updates $\tau_{opt}(1,s_6)$ to $0820$. Since all trip-transfers $\triptrans{t}{i}{t^\prime}{j}$ from trips $t_8$ and $t_{14}$ are such that $\triparr{t_1}{i}$ is greater than $\tau_{opt}(1,s_6)$, there are no other optimal journeys to $s_6$. Hence, we can safely remove $s_6$ from \textit{dlist}. Thus, as the algorithm proceeds, the destination list is progressively pruned, resulting in faster queries.  

Compared to the rTBTR algorithm, the One-To-Many version scans trip-segments fewer times. In the above example, repeated application of rTBTR scans the trip-segments $\tripsegment{t_1}{1}{3}$, $\tripsegment{t_2}{1}{3}$, $\tripsegment{t_8}{1}{3}$, and $\tripsegment{t_{14}}{1}{2}$ twice (once for each destination node) whereas the One-To-Many rTBTR does it only once.

The pseudocode for the One-To-Many rTBTR is presented in Algorithm \ref{alg:One-to-Many}. Lines 1--7 initialize the variables $\Lset(s)$, $\Jset(s)$, $\tripindexrange{t}$, and $\tau_{opt}(n,s)$. Line 8 iterates over all all possible departure times $\tau$ in a decreasing order. Lines 9--16 initialize $n$ and $Q_n$ as before. We start by copying all the elements of $\textit{dlist}$ into a dummy list \textit{dlist}$^\prime$ in Line 17. In the main \textit{while}-loop, for each allowed number of transfers $n\  (\leq\maxtrans)$, Line 19 first defines an empty set \textit{scope}. While scanning a trip-segment $\tripsegment{t}{h}{k}$, a \textit{for}-loop (Line 21) is introduced which iterates over all the stops in \textit{dlist}$^\prime$. Lines 22--32 scan the trip-segments as described in the TBTR algorithm. An additional step is introduced in Line 29 that adds stops whose labels could improve to \textit{scope}. Lastly, Lines 35--36 increment $n$ and update \textit{dlist}$^\prime$ using \textit{scope}.\\

\begin{breakablealgorithm}
\caption{One-To-Many rTBTR}
\label{alg:One-to-Many}

    \noindent\hspace*{\algorithmicindent}\textbf{Input}: GTFS, $\Tset$, $\source$, \textit{dlist}, $\maxtrans$, \textit{tlist}\\
    \hspace*{\algorithmicindent}\textbf{Output}: $\Jset$
    \begin{algorithmic}[1]
	\vspace{2mm}
		\State $\Lset(s),\Jset(s) \leftarrow \emptyset,\emptyset  \ \forall~s\in$ \textit{dlist} 
    \State $\tripindexrange{t} \leftarrow \triplen{t} \hspace{3pt}\forall$ $t \in T, \forall \, n=0, 1, \ldots,\maxtrans$ 
    \For{$\target \in$ \textit{dlist}}
            \State $\tau_{opt}(n,\target) \leftarrow \infty\  \forall \, n=0, 1, \ldots,\maxtrans$ 
	        \For {$s \in \neighbourhood{\target}$}
            \For { $(r,i)$ s.t. $\routestop=s$}
                \State $\Lset(\target)\leftarrow \Lset(\target) \cup \{(r,i,\footpath{s}{\target})\}$
        \EndFor
        \EndFor
        \EndFor

        \For{$\dep \in$ \textit{tlist}}        
	    \State $n\leftarrow 0$
		\State $Q_n \leftarrow \emptyset \ \forall \, n=0, 1, \ldots,\maxtrans$
        \State $clist \leftarrow \emptyset$ 
        \For {$s \in \neighbourhood{\source}$}
            \For { $(r,i)$ s.t. $\routestop=s$}
                \State $t\leftarrow$ earliest trip on $r$ s.t. $\dep + \footpath{\source}{s} \leq \tripdep{t}{i}$
                \State $clist \leftarrow clist \cup \big\{(t, i, n)\big\}$          
        \EndFor
        \EndFor
    \State \textsc{Enqueue}$(clist, \textit{ind}, Q_n)$
    \vspace{2mm}

\State \textit{dlist}$^\prime\leftarrow$ \textit{dlist}
        \While{$Q_n \neq \emptyset$ \AND $n\leq\maxtrans$}
            \State $scope\leftarrow \emptyset$
            \For {$\tripsegment{t}{h}{k} \in Q_n$}
                \For{$\target \in$ \textit{dlist}$^\prime$}
                    \For {$(\triproute{t}, i , \Delta\tau)\in\Lset(\target)$}
                    \If{$h<i\leq k$ \AND $\triparr{t}{i} + \Delta\tau<\tau_{opt}(n,\target)$}
                        \For{$m =n,n+1,\ldots\maxtrans$}
                        \If{$\tau_{opt}(m,\target)> \triparr{t}{i} + \Delta\tau$}
                        \State $\tau_{opt}(m,\target)\leftarrow \triparr{t}{i} + \Delta\tau$
                        \EndIf
                        \EndFor
                        \EndIf
                        \EndFor
                    \State $clist \leftarrow \emptyset$ 
                    \If{$\triparr{t}{h+1}<\tau_{opt}(n,\target)$}
                    \State add $s_d$ to $scope$ if not already present
                    \For {$\triptrans{t}{i}{t^\prime}{j} \in \Tset$ s.t. $h<i\leq k$ }
                    \State add $(t^\prime,j,n+1)$ to $clist$ if not already present
                    \EndFor
                    \EndIf
            \EndFor
            \State \textsc{Enqueue}$(clist, \textit{ind}, Q_{n+1})$
            \EndFor
        \For{$s \in$ \textit{dlist}}
            \State add $(\tau_{opt}(n,s), n)$ to $\Jset(s)$ if it is non-dominated

        \EndFor
    \State $n\leftarrow n + 1$
        \State \textit{dlist}$^\prime\leftarrow$ scope
        \EndWhile
        \EndFor

        \vspace{2mm}
        \setcounter{ALG@line}{0}
        \Procedure{\textsc{Enqueue}}{list of connections $clist$, vector \textit{ind}, queue $Q$}{
        \For{$(t,i,n) \in clist$}
            \If{ $i<\tripindexrange{t}$}
                \State $Q\leftarrow Q \cup \hspace{3pt}
               \big\{\tripsegment{t}{i}{\tripindex{n, t}}\big\}$
                \For { trip $t'$ s.t. $t \preceq t'$ \AND $\triproute{t}=\triproute{t'}$}
                \For {$j$ = $n,n+1,\ldots,\maxtrans$}
                    \State $ind(j,t')\leftarrow \min\big\{ind(j,t'), i\big\}$
                \EndFor
                \EndFor
    		\EndIf
        \EndFor
        }
        \EndProcedure

	\end{algorithmic}

\end{breakablealgorithm}

\textit{Proof of correctness:} Observe that if \textit{dlist}$^\prime$ was the same as \textit{dlist}, i.e., if the destination list was not pruned, then Algorithm \ref{alg:One-to-Many} is equivalent
to repeated application of rTBTR. Also, $s_d$ is deleted in some round $n<\maxtrans$ only when the \textit{if}-condition in Line 28 is violated, i.e., \say{for all trip-segments $\tripsegment{t}{h}{k}\in Q_n$, their first stops $t(h+1)$ are reached later than the best known arrival time at $s_d$}. Thus, to establish the algorithm's correctness, it is enough to show that the updates to $\tau_{opt}$ labels is identical in the following two cases: (i) \textit{dlist} is pruned (ii) \textit{dlist} is not pruned in round $n$ even when Line 28 is violated for all trip-segments. 

In Case (i), given that $s_d$ is removed in round $n$, $\tau_{opt}(m,s_d)$ will not be updated for $n+1\leq m \leq \maxtrans$ because the \textit{for}-loop in Lines 22--32 will not iterate over $s_d$. In Case (ii), for round $n+1$, we can show that there does not exist any $\tripsegment{t}{h}{k}\in Q_{n+1}$ that satisfies the arrival time condition in Line 23 and hence the algorithm will consequently not update $\tau_{opt}(n+1,s_d)$. To see why, note that for all trip segments $\tripsegment{t'}{i}{j}\in Q_n$, since Line 28 was violated, $\triparr{t'}{i+1}>\tau_{opt}(n,\target)$. Because every trip segment $\tripsegment{t}{h}{k} \in Q_{n+1}$ was added using some trip segment $\tripsegment{t'}{i}{j}\in Q_n$ in the previous round, we can conclude that $\triparr{t}{h} \geq \triparr{t'}{i+1} > \tau_{opt}(n,\target)$. At the start of round $n+1$, the value of $\tau_{opt}(n+1,\target)$ is same as $\tau_{opt}(n,\target)$ since it was set in Lines 24--26 in the previous round. Thus, $\triparr{t}{h} > \tau_{opt}(n,\target) = \tau_{opt}(n+1,\target)$ and the \textit{if}-condition in Line 23 is violated. Hence, the $\tau_{opt}$ label for round $n+1$ is not updated. A similar argument holds for later rounds.

\subsection{Multilevel extension: MHypTBTR}
\label{subsec:MHypTBTR}
\begin{figure}[htbp]
 \centerline{\includegraphics[scale=0.5]{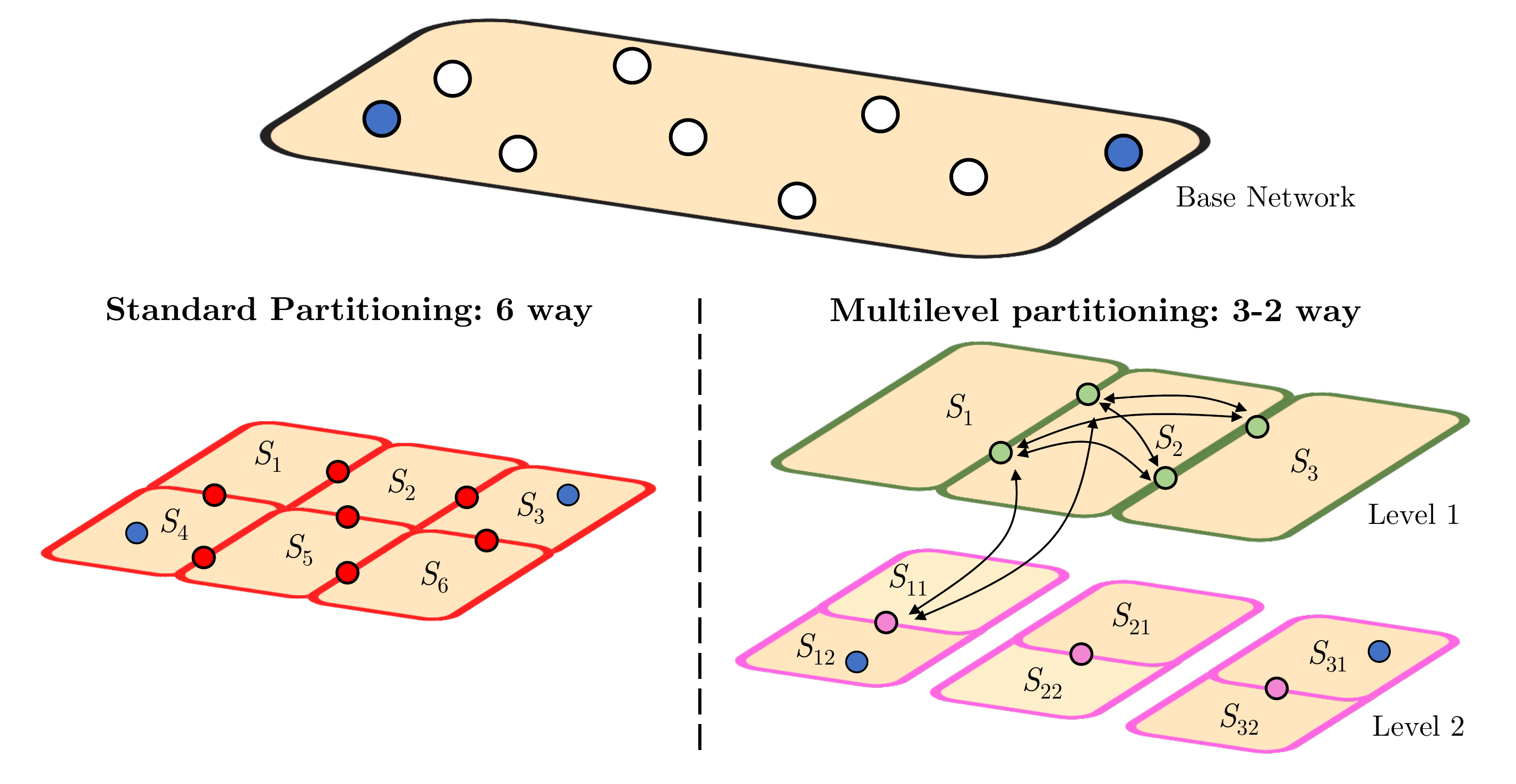}}
     \caption{Example to illustrate the advantages of multilevel partitioning}
     \label{fig:nested}
 \end{figure}

This section introduces MHypTBTR, i.e., HypTBTR combined with multilevel partitioning. While we propose this concept using HypTBTR, a similar approach can be used for HypRAPTOR. 
Consider the example in Figure \ref{fig:nested}. The base network is an abstraction of a transit network before partitioning. Nodes in white indicate stops, and those in blue show the source and destination. 

\begin{itemize}
\item Standard Partitioning: The figure on the left shows the network partitioned into six parts (as discussed in Section \ref{sec:reduce_querytime}). Red nodes represent cutstops. The labels show the stop cell ID for each partition $(S_1, S_2, \ldots, S_6)$ and $S_0$ is assumed to contain all the red cutstops. Thus, using standard partitioning, HypTBTR's fill-in trip set will require finding optimal journeys between ${7 \choose 2} 2!=42$ source-destination permutations of the cutsops.

\item Multilevel Partitioning: The figure on the right depicts multilevel partitioning with two levels. In Level 1, the network is partitioned into three \textit{parent partitions} $(S_1,S_2,S_3)$. Green nodes show the cutstops of Level 1. Next, in Level 2,  each parent partition is divided into two subparts (i.e., \textit{child partitions}). E.g.,  $S_1$ is divided into $S_{11}$ and $S_{12}$. Cutstops are shown using pink nodes here. Finding the fill-in set $\fillintrip$ can then be divided into two steps:
    \begin{enumerate}
        \item Compute the trips required to travel between cutstops at the topmost level, i.e., ${4 \choose 2} 2!=12$ source-destination permutations of the four green cutstops.
        \item For each parent partition, determine the trips required to travel between its cutstops and the cutstops of its children. E.g., in Figure \ref{fig:nested}, arrows between Levels 1 and 2 indicate 4 source-destination permutations for $S_1$ and its children $S_{11}$ and $S_{12}$. Similarly for $S_2$ and its children $S_{21}$ and $S_{22}$, we have 8 permutations. Thus, we get a total of $4+8+4=16$ permutations between the two levels.
    \end{enumerate}
    \end{itemize}
The overall number of source-destination pairs in the multilevel scheme is $12+16 =28$ compared to 42 in the standard scheme (a 33\% benefit). Note that if the parent partition is split into two parts, the optimal trips between the cutstops of sibling partitions are not required. However, if the split is performed differently, $\fillintrip$ would also include the optimal trips required to travel between siblings. 

The fill-in computation can be implemented in multiple ways. The most straightforward scheme is to store the fill-in trips in a single set $\fillintrip$, i.e., there is no distinction between trips required to travel between successive levels and within a level. In this case, MHypTBTR starts from $S_{12}$ (source stop cell) in Level 2 and travels up to Level 1 through a trip in $\fillintrip$. It then explores a much sparser graph and again transfers down to Level 2 using another trip in $\fillintrip$. A full Pareto-set is obtained since all the optimal trips between the cutstops are contained in $\fillintrip$. Alternately, a pointer for each trip indicating the level and cutstops for which it is optimal can be maintained (similar to how multilevel arc-flags are used in road networks). This allows additional pruning because, for every trip in $\fillintrip$, we have some extra information on the source-destination pairs for which the trip was optimal. However, the differences between these methods are likely to be pronounced in the query phase. Since the main goal here is to decrease preprocessing, we adopt the former fill-in scheme due to its simplicity.

\section{Experiments}
\label{sec:Experiments}
All algorithms used in our experiments were implemented in Python 3, and the source codes are available at \href{https://github.com/transnetlab/transit-routing}{github.com/transnetlab/transit-routing}. The query phase codes were run on an Intel Core i7-8700 CPU clocked at 3.2 GHz with 32 GB RAM. The more intensive algorithms related to preprocessing were evaluated in parallel using a 128-core Intel Xeon Gold CPU clocked at 3.0 GHz with 512 GB RAM. The metrics obtained using the parallel method have been marked with an asterisk (\textit{$\Tset$-time}$^*$ and \textit{$\fillintrip$-time}$^*$). Query times were averaged over a set of 10\smp000 randomly selected source and destination stops. For comparison, we also include results from Dijkstra's algorithm on a Time-Expanded graph (labeled as TED). A maximum transfer limit of four was used for all the algorithms (except TED). 

We tested our algorithms on six transit networks: Switzerland, Netherlands, Sweden, Israel, Taichung, and Bangalore. Most of these are open data sets (Source: \href{https://transitfeeds.com/}{transitfeeds.com}) except for Bangalore. Bangalore's network is derived from the Bangalore Metropolitan Transport Corporation (BMTC) timetable, which is the sole public transit agency that operates buses in the city. Note that while the first four networks are country-level networks that combine multiple modes such as buses, rails, trams, and metro, the last two datasets correspond to city-level networks. Table \ref{tab:network_des} summarizes various statistics for these networks for a single day (pre-COVID-19). Columns \textit{hypedges} and \textit{hypnodes} represent the number of hyperedges and hypernodes in the corresponding hypergraph, respectively. A significant amount of preprocessing was required to generate these datasets. For example, country-level datasets often contain overtaking trips since they integrate timetables from multiple agencies and modes. As the algorithms discussed in the present study require trips to follow the FIFO property, overtaking trips were discarded. Also, none of the above-mentioned datasets provide footpath details. This information was separately extracted from OpenStreetMaps (OSM) \href{https://download.geofabrik.de/}{(geofabrik.de)}. To do so, all stops were snapped to the nearest OSM coordinate and the corresponding distance matrix was calculated. Next, assuming a constant walking speed of 1 m/s, we obtained footpath times. Using a threshold on walking time, these footpath edges are then filtered and subsequently adjusted to satisfy transitivity and triangle inequality. The values in parenthesis in the \textit{footpaths} column of Table \ref{tab:network_des} indicates the threshold on walking time in seconds. Note that these values represent the initial threshold used in the OSM network but since more footpaths are added to ensure transitivity, the final footpath graph can contain edges which take longer to walk.

\newcolumntype{g}{>{\columncolor{Gray}}r}
\begin{table}[h]
    \caption{Network statistics (\textit{stops}: number of stops, \textit{routes}: number of routes, \textit{trips}: number of trips, \textit{footpaths}: number of footpaths and threshold on walking time in seconds, \textit{stopevents}: total stop events, \textit{hypedges}: number of of hyperedges, \textit{hypnodes}: number of hypernodes) \vspace{-5mm}}
    \begin{center}
    \begin{tabular}{r g r g r g r g }
    \hline
    \textbf{Network} & \textbf{stops} & \textbf{routes}& \textbf{trips}& \textbf{footpaths}& \textbf{stopevents}& \textbf{hypedges}& \textbf{hypnodes}\\
    \hline
    Switzerland & 22\smp885 & 9\smp094 & 146\smp228 & 87\smp820 (180)& 1\smp403\smp398 & 11\smp753 & 53\smp004\\
    Netherlands & 41\smp151 & 5\smp598  & 87\smp924 & 69\smp832 (180) & 1\smp697\smp456 & 38\smp047  & 40\smp514\\
    Sweden & 34\smp464 & 10\smp408 & 46\smp794 & 133\smp024 (180) & 899\smp410 & 14\smp195 &76\smp920\\
    Israel & 25\smp993 & 5\smp665 & 90\smp050 & 122\smp134 (120) & 2\smp813\smp973 & 22\smp719  & 66\smp732\\
    Taichung & 8\smp261 & 438 & 10\smp052 & 48\smp834 (120) & 539\smp271 & 7\smp732  & 24\smp855\\
    Bangalore & 8\smp323 & 5\smp590 & 27\smp202 & 508\smp998 (120) & 853\smp153 & 8\smp203  & 260\smp089\\
    \hline
\end{tabular}
\label{tab:network_des}
\end{center}
\end{table}

Table \ref{tab:level1query} presents the query performance (in milliseconds) of the non-partitioning algorithms. The following list summarizes our findings.
\vspace{-3mm}
\begin{itemize}
    \item TED vs. others:  The query times reported are from \href{https://networkx.org/}{NetworkX}'s implementation of the Dijkstra's algorithm on a time-expanded graph (TED). The results reconfirm that modern PTR algorithms such as RAPTOR and TBTR perform better than conventional Dijkstra-based approaches.
    
    \item RAPTOR vs. TBTR: TBTR outperforms RAPTOR in all three instances mainly because it uses the trip-transfers set $\Tset$ to directly switch between trips as opposed to RAPTOR's method of finding the earliest trip to board a route. Results from the preprocessing phase of TBTR are in Appendix \ref{sec:tbtr_preprocessing}.
    \item rTBTR vs. One-To-Many rTBTR: To compare rTBTR against its One-To-Many variant, we first randomly selected 70 source stops. For each of these source stops, 70 random destinations were chosen, and a profile query was run. That is, rTBTR queries were run 4\smp900 times, whereas the OTM rTBTR was run 70 times with different source stops as input. In all the test cases, our implementation significantly beats repeated application of rTBTR by 90--95\%, demonstrating the potential of the One-To-Many version in reducing the preprocessing time of HypTBTR. 
    \item rRAPTOR vs. One-To-Many rRAPTOR: rRAPTOR was also upgraded to handle One-To-Many queries (see Appendix \ref{sec:raptor_appen} for pseudocodes), similar to One-To-Many rTBTR, by pruning the destination list for faster queries. The experimental setup used to compare rRAPTOR and One-To-Many rRAPTOR is the same as described above. One-To-Many rRAPTOR was also found to improve runtimes by 98\% when compared with repeated application of rRAPTOR.
\end{itemize}

\newcolumntype{g}{>{\columncolor{Gray}}c}
\begin{table}[H]
    \caption{Query times (in milliseconds) of various PTR algorithms (TED: Time-expanded Dijkstra, RAPTOR and rRAPTOR: \cite{delling2015round}, TBTR and rTBTR: \cite{witt2015trip}, OTM (One-To-Many) rTBTR: Algorithm \ref{alg:One-to-Many}, OTM (One-To-Many) rRAPTOR: Algorithm \ref{alg:One-To-Many_rRAPTOR})\vspace{-5mm}} 
    \begin{center}
    \begin{tabular}{c g c g c g c g}
    \hline
    \textbf{Network}  & \textbf{TED} & \makecell{\textbf{RAPTOR}} & \makecell{\textbf{TBTR} } & \makecell{\textbf{rTBTR}}  & \makecell{\textbf{OTM}\\ \textbf{rTBTR} } & \makecell{\textbf{rRAPTOR}}  & \makecell{\textbf{OTM}\\ \textbf{rRAPTOR}}\\
    \hline
    Switzerland &   12\smp847.5 &   334.7  & 96.9 & 81\smp710.6 & 3\smp654.1 & 413\smp445.3 &7\smp657.1\\
    Netherlands & 22\smp738.8   &  155.1   & 43.1 & 19\smp952.4 & 1\smp192.2 & 196\smp646.5 &3\smp131.2\\
    Sweden &   1\smp568.8 &   103.5  & 6.3 & 1\smp250.9 & 131.8 & 45\smp668.7 & 685.5\\
    \hline
    \end{tabular}
\label{tab:level1query}
    \end{center}
\end{table}

\begin{figure}[t]
 \centerline{\includegraphics[scale=0.55]{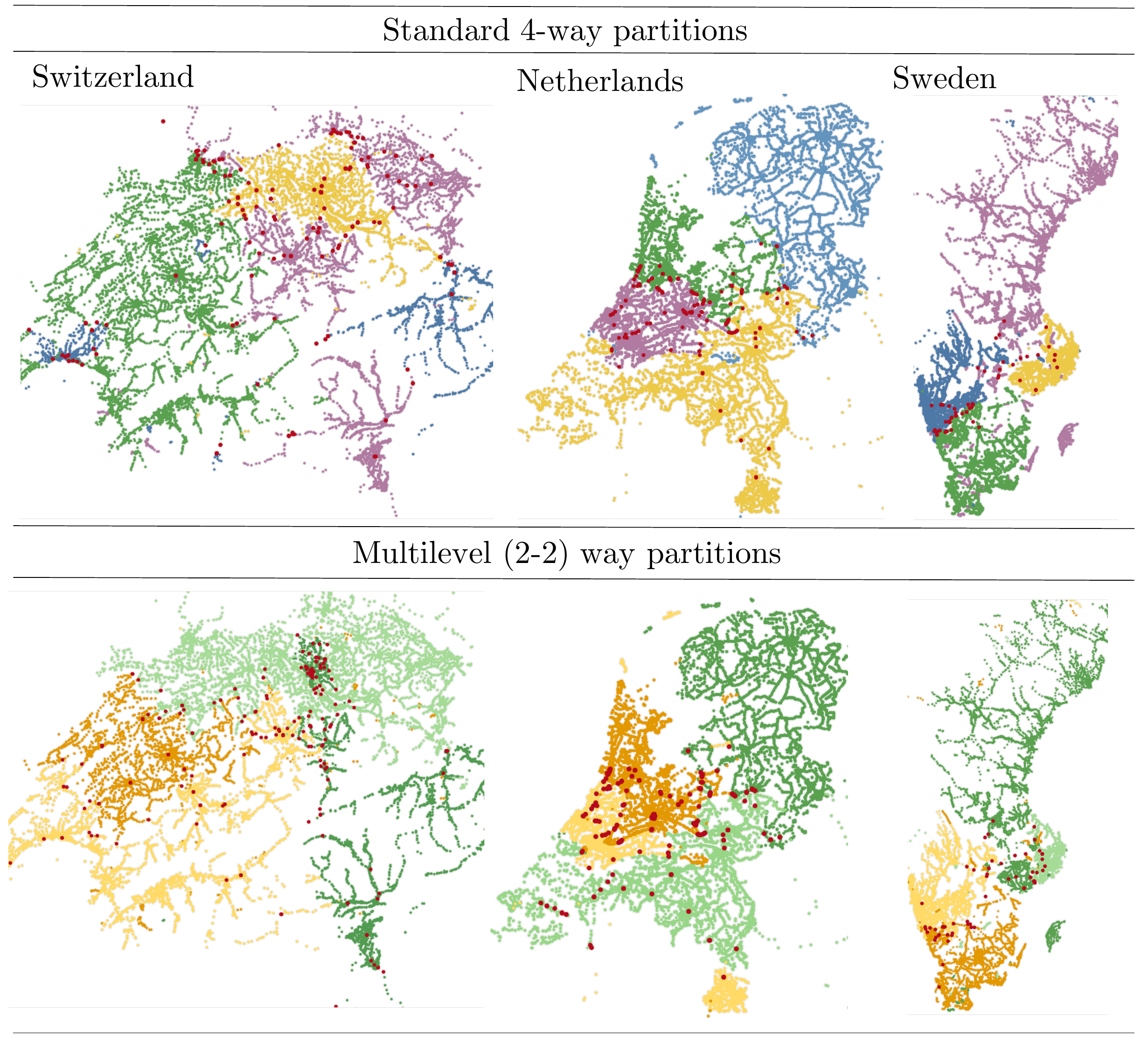}}
     \caption{Illustration of standard 4-way and multilevel (2-2) partitioning (Partitions in standard partitioning are indicated by blue, green, yellow, and purple. For the multilevel (2-2) way case, parent partitions are in green and yellow. Child partitions are shown using dark and light shades, respectively. Red dots indicate cutstops.)}
     \label{fig:network_part}
 \end{figure}

Next, we analyzed the preprocessing phase of HypTBTR (see Table \ref{tab:hyptbtr_pre_results}). The trip-transfer computation phase of HypTBTR is the same as that of TBTR. For the fill-in computation phase, we first generate a hypergraph using the GTFS data. Table \ref{tab:network_des} contains the details of these hypergraphs. The parameters used for the KaHyPar algorithm are \textit{seed} -1, \textit{epsilon} 0.2, and \textit{cut\_kKaHyPar\_sea20.ini} configuration. The top panel in Figure \ref{fig:network_part} shows the stops when the test networks are partitioned into four cells (blue, green, yellow, and purple) using the standard method. The multilevel partitions are shown in the bottom panel, where shades of the same color depict siblings of a parent partition. Cutstops are indicated in red. The Sc$_3$ weighting scheme was used in these experiments, i.e., the weight of a hyperedge corresponding to a stop $s$ is set to the logarithm of total stopevents associated with the stops in $\neighbourhood{s}$. The weight of a hypernode is determined by the total number of events along the corresponding route. A comparison of the weighting schemes discussed in Section \ref{subsec:HypTBTR} is presented in Appendix \ref{sec:weighting_scheme_compari}. The parameters reported in Table \ref{tab:hyptbtr_pre_results} include: \textit{scut}: number of cutstops and \% w.r.t total stops in the network, \textit{pqueries}: number of profile queries required to compute the fill-in set, \textit{$\fillintrip$ size}: \% of trips that are part of fill-in, and \textit{$\fillintrip$ time}: time for computing fill-in trips. For each of the three networks, results from both standard and multilevel partitioning are reported. A column label with label 10 (5-2) implies that the standard partitioning approach splits the network into ten parts, and the multilevel partitioning method creates five parent partitions at the upper level and each parent is divided into two child partitions. The following observations are noteworthy. 
\begin{itemize}
    \item With an increase in the number of partitions, both \textit{scut} and \textit{pqueries} increase as expected.
    \item The values in the parenthesis indicate the \% benefit of the multilevel partition over its standard version. For example, in Sweden's ten partitioning case, the number of pqueries needed are 46\smp440 (standard) and 19\smp362 (multilevel), which translates to a benefit of 58.3\%. A significant reduction can be seen in terms of the \textit{pqueries} and \textit{$\fillintrip$ time} across standard and multilevel versions in all test cases. An advantage of comparing \textit{pqueries} is that it is a language-agnostic metric. The benefits are mostly in the range of 5--58\%. 

    \item While multilevel partitioning was consistently better in all test cases, the percentage benefit varies. This is mainly because the partitions (and hence the cutstops) generated depend on the initial KaHyPar configuration such as \textit{seed} and \textit{epsilon}. However, since the aim here is to compare the benefits of multilevel partitioning over standard partitioning, the configuration parameters used were kept same in all the experiments.   
\end{itemize}

\begin{table}[t]
    \caption[HypTBTR preprocessing using KaHyPar]{HypTBTR preprocessing using KaHyPar (\textit{scut}: cutstop count and \% of cutstops, \textit{pqueries}: profile queries required, \textit{$\fillintrip$ size}: \% of fill-in trips, \textit{$\fillintrip$ time}$^*$: time in seconds to compute $\fillintrip$. Values in teal indicate \% gain in the multilevel version over its standard counterpart.) \vspace{-5mm}}
     \begin{center}
    \begin{tabular}{c c c c g c g}
    \hline
    \multirow{2}{*}{\textbf{Network}} & \multirow{2}{*}{\textbf{Partitioning}}&
    \multirow{2}{*}{\textbf{Metric}} & \multicolumn{4}{c}{\textbf{Partitions}}\\
    &&& \textbf{4 (2-2)}& \textbf{6 (3-2)}& \textbf{8 (4-2)}& \textbf{10 (5-2)}\\
    \hline
    \multirow{8}{*}{Switzerland} & \multirow{4}{*}{standard}& \textit{scut}     &  263 (1.1\%)    &  374   (1.6\%)    &   377 (1.6\%)&  437 (1.9\%)\\
    & &\textit{pqueries} &     68\smp906 &  139\smp502  & 141\smp752 & 190\smp532 \\
    & &\textit{$\fillintrip$ size}   & 10.7\%      & 18\%    & 17.4\% & 18.2\%\\
    & &\textit{$\fillintrip$ time}$^*$   &  256.2    &   484.6  &  492.1 & 655.5 \\[.15cm]
    \cdashline{3-7}
     \rule{0pt}{.45cm} & \multirow{4}{*}{multilevel} & \textit{scut}   &    273 (1.2\%) &   356 (1.6\%) & 396 (1.7\%)&  539  (2.4\%)\\
    & &\textit{pqueries}   &   55\smp485 \textcolor{teal}{(19.5\%)} &    98\smp705 \textcolor{teal}{(29.2\%)}& 113\smp131 \textcolor{teal}{(20.2\%)}& 184\smp528 \textcolor{teal}{(3.2\%)}\\
    & &\textit{$\fillintrip$ size}   & 12.6\%      & 17.3\%    & 18.8\% & 21.2\%\\
    & & \textit{$\fillintrip$ time}$^*$   &   214.1 \textcolor{teal}{(16.4\%)}  &    334.8 \textcolor{teal}{(30.9\%)}& 386.3 \textcolor{teal}{(21.5\%)}& 594.5 \textcolor{teal}{(9.3\%)}\\
    \hline
    \multirow{8}{*}{Netherlands} &  \multirow{4}{*}{standard} & \textit{scut}     &  273 (0.7\%)    &   329 (0.8\%)    &  483 (1.2\%)&  567 (1.4\%)\\
    & &\textit{pqueries} &  74\smp256     & 107\smp912   & 232\smp806 & 320\smp922 \\
    & &\textit{$\fillintrip$ size}   & 20.9\%      & 28.3\%    & 37.5\% & 39.2\%\\
    & &\textit{$\fillintrip$ time}$^*$   & 205.7     & 274.2    & 494.9 & 714.3 \\[.15cm]
    \cdashline{3-7}
    \rule{0pt}{.45cm} &  \multirow{4}{*}{multilevel} &\textit{scut}   &   288 (0.7\%)  & 368 (0.9\%)   & 516 (1.2\%) & 613 (1.5\%)\\
    & &\textit{pqueries}   &     63\smp063 \textcolor{teal}{(15.1\%)}& 97\smp180    \textcolor{teal}{(9.9\%)}& 152\smp821 \textcolor{teal}{(34.4\%)}& 187\smp971 \textcolor{teal}{(41.4\%)}\\
    & &\textit{$\fillintrip$ size}   & 18.3\%      & 27\%    & 33.4\% & 41.3\%\\
    & & \textit{$\fillintrip$ time}$^*$   &  191.4  \textcolor{teal}{(7.0\%)}  & 242.9  \textcolor{teal}{(11.4\%)}  & 473.1 \textcolor{teal}{(4.4\%)}& 489.6 
    \textcolor{teal}{(31.5\%)}\\
    \hline
\multirow{8}{*}{Sweden} &  \multirow{4}{*}{standard} & \textit{scut}     &  77  (0.2\%)     &    123  (0.4\%) &  163 (0.5\%)&  216 (0.6\%)\\
     & &\textit{pqueries} & 5\smp852      &15\smp006    & 26\smp406 & 46\smp440\\
     & &\textit{$\fillintrip$ size}   & 6.4\%      & 14\%    & 13.8\% & 19.7\%\\
     & &\textit{$\fillintrip$ time}$^*$   & 9.1     & 12.9    & 17.4 & 22.6\\[.15cm]
    \cdashline{3-7}
     \rule{0pt}{.45cm} & \multirow{4}{*}{multilevel} &\textit{scut}   &    87 (0.2\%)  & 137 (0.4\%)    & 162 (0.5\%) & 204  (0.6\%)\\
     & &\textit{pqueries}   &    2\smp808 \textcolor{teal}{(52\%)} & 6\smp922  \textcolor{teal}{(53.9\%)}  & 12\smp729 \textcolor{teal}{(51.8\%)}& 19\smp362 \textcolor{teal}{(58.3\%)}\\
     & &\textit{$\fillintrip$ size}   & 2.4\%      & 9.7\%    & 11.2\% & 15.7\%\\
     & & \textit{$\fillintrip$ time}$^*$   &  4.4 \textcolor{teal}{(51.6\%)}   &   7.1 \textcolor{teal}{(45.0\%)} & 9.5 \textcolor{teal}{(45.4\%)}& 10.6 \textcolor{teal}{(53.1\%)}\\
    \hline
   \end{tabular}
\label{tab:hyptbtr_pre_results}
    \end{center}
\end{table}

Table \ref{tab:partitiong_query_results} contains the query time (in milliseconds) for HypRAPTOR, HypTBTR, and their multilevel versions. Based on these results, the following conclusions can be drawn.
\begin{itemize}
    \item HypRAPTOR vs. RAPTOR:  As expected, HypRAPTOR performs better than RAPTOR in all the test cases. The benefits were found to be in the range of 12--32\%.

    \item HypTBTR vs. TBTR: HypTBTR consistently performs better than TBTR on all networks. The gain observed was in the range of 23--37\%.  

    \item HypRAPTOR vs. HypTBTR: In all the three networks, we observe that the average gains from using HypTBTR over TBTR are more than that between HypRAPTOR and RAPTOR. In other words, the partitioning-based scheme performed better in the TBTR setting than RAPTOR. A possible reason could again be that TBTR uses trips as building blocks instead of routes. For example, imagine a route with $m$ trips of which only one is part of the fill-in set. While HypRAPTOR adds the route (and all $m$ trips) to its fill-in set, HypTBTR adds only one trip. 

    \item MHypTBTR vs. HypTBTR and MHypRAPTOR vs. HypRAPTOR: Query times of the multilevel versions are in the same range as their standard counterparts. This is expected since the size of the fill-in trips in multilevel partitioning is approximately the same as that of standard partitioning. Multilevel partitioning mainly helped reduce preprocessing times.
\end{itemize}

\newcolumntype{g}{>{\columncolor{Gray}}r}

\begin{table}[H]
\renewcommand{\arraystretch}{1.2}
    \caption{Query performance (in milliseconds) of algorithms with partitioning-based speed-up (Values in teal indicate \% gain over their base variant.) \vspace{-5mm}}
    \begin{center}
    \begin{tabular}{r r g r g r}
    \hline
    \multirow{2}{*}{\textbf{Network}} & 
    \multirow{2}{*}{\textbf{Metric}} & \multicolumn{4}{c}{\textbf{Partitions}}\\
    && \textbf{4 (2-2)}& \textbf{6 (3-2)}& \textbf{8 (4-2)}& \textbf{10 (5-2)}\\
    \hline
    \multirow{4}{*}{Switzerland} & HypRAPTOR & 230.6 \textcolor{teal}{(31.1\%)} & 240.1 \textcolor{teal}{(28.3\%)}& 231.4 \textcolor{teal}{(30.9\%)} & 226.7 \textcolor{teal}{(32.3\%)} \\

     & HypTBTR &    
     62.4 \textcolor{teal}{(35.6\%)}& 
     67.3 \textcolor{teal}{(30.5\%)}& 
     63.4 \textcolor{teal}{(34.6\%)}& 
     61.4 \textcolor{teal}{(36.6\%)}\\

     & MHypRAPTOR & 230.2 \textcolor{teal}{(31.2\%)}& 
                    231.9 \textcolor{teal}{(30.7\%)}& 
                    232.4 \textcolor{teal}{(30.6\%)}& 
                    235.8 \textcolor{teal}{(29.5\%)}
                         \\

     & MHypTBTR &   
     65.0 \textcolor{teal}{(32.9\%)} & 
     65.1 \textcolor{teal}{(32.8\%)}&  
     66.2 \textcolor{teal}{(31.7\%)} &  
     65.4 \textcolor{teal}{(32.5\%)} \\
    \hline
    \multirow{4}{*}{Netherlands} & HypRAPTOR & 117.9 \textcolor{teal}{(22.6\%)} & 118.6 \textcolor{teal}{(21.3\%)}& 119.4  \textcolor{teal}{(22\%)} & 123.6 \textcolor{teal}{(21.5\%)} \\

    &    HypTBTR &  28.2  \textcolor{teal}{(32.7\%)}& 28.9
                             \textcolor{teal}{(33.7\%)}& 29.9
                             \textcolor{teal}{(31.7\%)}& 29.6
                             \textcolor{teal}{(29.6\%)}\\

    &    MHypRAPTOR & 127.3 \textcolor{teal}{(22.9\%)} & 126.8  \textcolor{teal}{(22.8\%)}& 
    127.1 \textcolor{teal}{(23.4\%)} & 
    127.2 \textcolor{teal}{(23.5\%)} \\

    &    MHypTBTR &  28.3  \textcolor{teal}{(33.7\%)}& 28.4
                             \textcolor{teal}{(33.4\%)}& 28.9
                             \textcolor{teal}{(32.2\%)}& 28.6
                             \textcolor{teal}{(33.7\%)}\\
    \hline
    \multirow{4}{*}{Sweden} & HypRAPTOR & 90.9 \textcolor{teal}{(12.1\%)} &  88.7 \textcolor{teal}{(14\%)}& 89.6  \textcolor{teal}{(13.4\%)} & 91.1 \textcolor{teal}{(12\%)} \\

     & HypTBTR &   4.8 \textcolor{teal}{(23.8\%)}
     & 4.5 \textcolor{teal}{(28.6\%)}
     & 4.4 \textcolor{teal}{(30.2\%)}
     & 4.3 \textcolor{teal}{(31.7\%)}\\

     & MHypRAPTOR & 93.5 \textcolor{teal}{(9.7\%)} &  
     93.4 \textcolor{teal}{(9.7\%)}&  
     85.2 \textcolor{teal}{(17.7\%)} &  
     85.1 \textcolor{teal}{(17.8\%)} \\
 
     &  MHypTBTR 
     &  4.7  \textcolor{teal}{(25.4\%)}
     & 4.3 \textcolor{teal}{(31.7\%)}
     & 4.4   \textcolor{teal}{(30.2\%)}
     & 4.3      \textcolor{teal}{(31.7\%)}\\
    \hline
     \end{tabular}
    \label{tab:partitiong_query_results}
    \end{center}
\end{table}

\begin{figure}[h]
 \centerline{\includegraphics[scale=0.55]{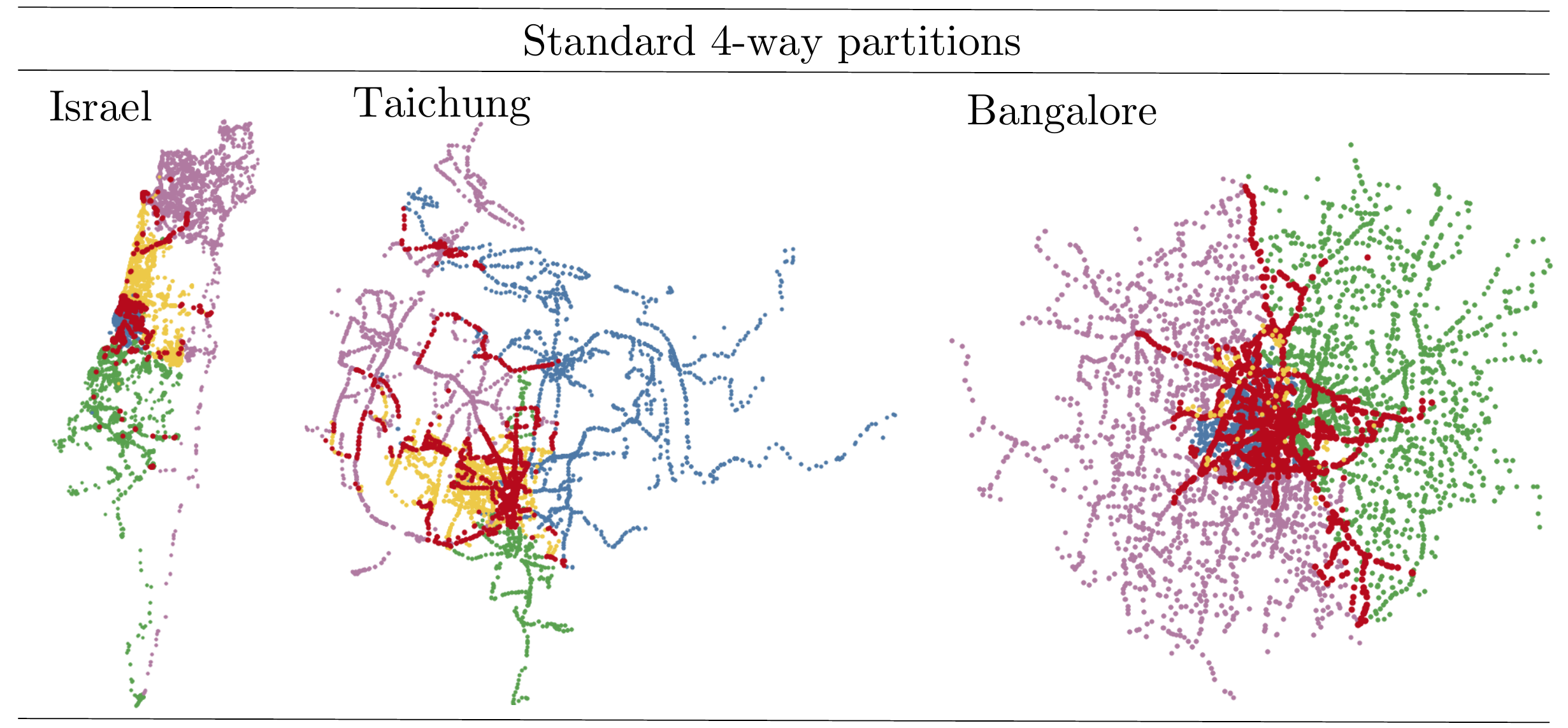}}
     \caption{Standard 4-way partitioning of Israel, Taichung, and Bangalore (Partitions in standard partitioning are indicated by blue, green, yellow, and purple. Red dots indicate cutstops.)}
     \label{fig:wronng_network_part}
 \end{figure}

While the benefits in query times of the partitioning-based methods depend on many factors such as the partitioning algorithm and weighting scheme used, the network topology also plays a key role. While experimenting with different networks, it was observed that the proposed methods do not always perform well. To understand why, recall that the benefits in query times are inversely related to the size of the fill-in set (since the network explored during the query phase comprises of fill-in and source/destination regions). For some transit networks such as Israel, Taichung, and Bangalore, the fill-in set size was found to be relatively high. Table \ref{tab:NetworkMetrics} highlights a few metrics that distinguish these networks. Columns \textit{spr} and \textit{eps} denote the average number of routes and events (arrival/departure) per stop. KaHyPar was used for partitioning and the results reported are for four partitions. Figure \ref{fig:wronng_network_part} shows the locations of cutstops for these networks.

\newcolumntype{g}{>{\columncolor{Gray}}c}

\begin{table}[h]
    \begin{center}
    \caption{Comparison of network metrics (\textit{scut}: cutstop count and \% of cutstops, \textit{TBTR q-time}: Average TBTR query time in milliseconds,
    \textit{HypTBTR q-time}: Average HypTBTR query time (in milliseconds) and \% improvement over TBTR, \textit{$\fillintrip$ size}: \% of fill-in trips, 
    \textit{spr}: average stops per route, \textit{eps}: average events per stop), $\Tset$ \textit{size}: size of trip-transfer set in millions)}
    \begin{tabular}{c r g r g c g c}
    \hline
    \textbf{Network} & \textbf{scut} &  \makecell{\textbf{TBTR}\\\textbf{q-time}} & \makecell{\textbf{HypTBTR}\\\textbf{q-time}}& \textbf{$\fillintrip$ size}&  \textbf{spr} & \textbf{eps}& $\Tset$ \textit{size}\\
    \hline
    Switzerland & 263 (1.1\%) & 96.9 & 62.4 \textcolor{teal}{(35.6\%)} & 10.7\% & 16.7 & 61.3  & 2.4\\
    Netherlands & 273 (0.7\%) & 43.1 & 28.2 \textcolor{teal}{(32.7\%)} &  20.9\%& 19.9 & 41.2  & 2.1\\
    Sweden & 77 (0.2\%) & 6.3 & 4.8 \textcolor{teal}{(23.8\%)} & 6.4\% & 24.2 & 26.1 & 0.7\\
    Israel & 936 (3.6\%) & 125.2 & 127.0 \textcolor{teal}{(-1.4\%)}& 40.9\% &35.9 & 108.3 & 7.5\\
    Taichung & 1\smp309 (15.8\%) & 56.4& 59.6 \textcolor{teal}{(-5.7\%)} & 95.2\% & 53.4 & 65.3 & 1.4\\
    Bangalore & 1\smp713 (20.6\%) & 231.3 & 278.4 \textcolor{teal}{(-20.3\%)} & 98.9\% & 35.6  & 102.5 & 14.1\\
    \hline
    \end{tabular}
\label{tab:NetworkMetrics}
    \end{center}
\end{table}

As can be seen from the table and the figure, the \% of cutstops in Israel, Taichung, and Bangalore is significantly higher compared to the ones studied earlier. As a result, the size of fill-in trips is also drastically higher. While we can influence the partitions to some extent using different weighting schemes and partitioning algorithms, our empirical tests indicated that the \% of cutstops mainly depended on the network topology, particularly the density of footpath graph and the extent of overlapping routes (represented by \textit{spr} and \textit{eps}). For this reason, we even lowered the initial walking threshold (Table \ref{tab:network_des} Column \textit{footpaths}) from 180 to 120 s for the latter three networks. Since footpaths are joined to make the graph complete under transitivity, a larger threshold in a dense network like Bangalore might form a full clique resulting in unrestricted walking. As the number of partitions increase, the number of cutstops (and hence the size of the fill-in set) increases. Also, recall that in multilevel partitioning, the size of the fill-in set was found to be close to that of the corresponding standard partitioning case. Multilevel partitioning mainly speeds-up the fill-in computation phase. Thus, we did not notice significant benefits from creating more partitions or from multilevel partitioning for Israel, Taichung, and Bangalore.

\section{Conclusions}
\label{sec:Conclusions}
Seamless journey planning plays an essential role in improving the attractiveness of public transit. Given the widespread use of mobile applications for this purpose, the algorithms used in the backend must be fast and efficient. To achieve this goal, we modified an existing well-known Trip-based Transit Routing (TBTR) algorithm for efficiently solving One-To-One queries by extending it to a graph partition-based method called HypTBTR. A new weighting scheme for partitioning the hypergraph was introduced. The benefits observed in the query times were in the range of 23--37\% on three country-level open datasets. Since query time reduction comes at the expense of increased preprocessing, we introduced a One-To-Many version for range queries. 

The proposed extension not only improves preprocessing times by solving One-To-Many profile queries 90--95\% faster, but also makes the TBTR approach more practical. This is because users often query for the shortest path between two locations, and a location can have multiple stops near it. Furthermore, we also explored a multilevel partitioning paradigm compatible with HypTBTR and HypRAPTOR, which further reduces the fill-in computation time by 5--53\%. The proposed extensions enable TBTR to handle queries in large-scale networks by eliminating preprocessing-related bottlenecks, making it more scalable. Our experiments also revealed instances where the number of cutstops generated by partitioning algorithms during preprocessing were prohibitively large. One could explore other objectives and weighting schemes in the partitioning algorithms to tackle these instances. Partitions generated from past source-destination query data can also make the proposed algorithms more practical for mobile and real-time applications. 

Other algorithms such as CSA and ACSA were not used for comparison as they only evaluate the earliest arrival time query. Also, Transfer Patterns-based algorithms (Transfer Patterns, and Scalable Transfer Patterns) are expected to perform better, but the preprocessing associated with them is likely to be much higher than algorithms in the current study. 
We defer these comparisons and multilevel partitioning extensions of Transfer Patterns for future research. Apart from partitioning, the goal-directed method is another successful technique that is generally studied in road routing for faster One-To-One queries. E.g., \cite{finkelstein:hal-03010079} extended the CSA to Goal-Directed CSA. Similar extensions can be explored for RAPTOR and TBTR. 

Most results in recent transit routing literature, including ours, are empirical in nature. In PTR, for the Earliest Arrival Problem, we can model the transit timetable as a TE-graph and run Dijkstra's algorithm to get the fastest route. One can thus argue that the Earliest Arrival Problem can be solved in polynomial time (as a function of the number of nodes and edges in the TE-graph) \citep{kaufman1993fastest, chabini1998discrete}. Multiobjective problems, on the other hand, are in general NP-Hard since the efficiency frontier can have an exponential number of solutions (see \cite{tarapata2007selected} for more details). In special cases where one of the criteria is the number of transfers with a maximum allowable transfer limit, one can create a finite number of layers of the TE-graph and use the Dijkstra's method to compute the Pareto set \citep{brodal2004time}. Hence, this PTR variant can also be solved in polynomial time. Note that the above discussion is valid only for networks with FIFO property. 

\cite{delling2015round} provided a loose bound of $\mathcal{O}(\maxtrans(\Sigma_{r\in R}(\routelen{r})+|T|+|F|))$ on the worst-case complexity of RAPTOR using their naive variant (i.e., assuming the local and destination-based pruning are not used). In such a setting, every route is scanned exactly once per round. However, the above expression does not reflect the true complexity of the RAPTOR algorithm because its success comes mainly from the marking and pruning methods. TBTR's pruning methods are more involved, because of which a worst-case complexity result is difficult to derive. Computing optimal partitions for preprocessing on the other hand is known to be NP-Hard \citep{lengauer2012combinatorial}. In general, research on complexity analysis is sparse and is another open topic for future research.

\textbf{Acknowledgments.} The authors would like to thank the Managing Director of BMTC for facilitating data sharing. Special thanks to Mr. Dibya Ranjan Jena, Consultant (BMTC) for helping us understand BMTC's operations, GPS, and ticketing data. The authors also appreciate Hemant Gehlot's comments on an earlier draft of this article. The first author would also like to thank John Varghese George, Saumya Bhatnagar, and Nipun Choubey for useful discussions on programming and version control.

\newpage
\bibliography{figures/reference} 
\clearpage

\appendix
\section{Appendix: Preprocessing phase of the TBTR algorithm}
\input{appendix_A}

\newpage
\section{Appendix: Extensions of RAPTOR}
\input{appendix_B}
\newpage
\section{Appendix: Effect of weighting schemes on hypergraph partitioning}
\input{appendix_C}

\newpage
\section{Appendix: Performance of alternate hypergraph partitioning methods}
\input{appendix_D}

\end{document}

%% file: figures/symbol_list.tex
\newcommand{\source}{{s_o}}
\newcommand{\target}{{s_d}}
\newcommand{\footpath}[2]{{f(#1, #2)}}
\newcommand{\footstops}[2]{{(#1, #2)}}

\newcommand{\prearr}[1]{{\tau_{arr}(#1)}}

\newcommand{\originstopcell}{{S_o}}
\newcommand{\targetstopcell}{{S_d}}
\newcommand{\originroutecell}{{R_o}}
\newcommand{\targetroutecell}{{R_d}}

\newcommand{\tripflag}[1]{{flag(#1)}}
\newcommand{\tripstop}[2]{{#1(#2)}}
\newcommand{\tripindex}[1]{{ind(#1)}}
\newcommand{\tripindexrange}[1]{{ind(n,#1)}}
\newcommand{\triproute}[1]{{route(#1)}}
\newcommand{\triparr}[2]{{arr(#1, #2)}}
\newcommand{\tripdep}[2]{{dep(#1, #2)}}
\newcommand{\tripsegment}[3]{{#1(#2\rightarrow#3)}}
\newcommand{\triptrans}[4]{{\big(#1,#2, #3 , #4\big)}}
\newcommand{\triplen}[1]{{l_{#1}}}

\newcommand{\dep}{{\tau}}
\newcommand{\maxtrans}{{\lambda}}

\newcommand{\route}{r}
\newcommand{\routestop}{{r(i)}}
\newcommand{\routelen}[1]{{l_{#1}}}

\newcommand{\fillintrip}{{\mathcal{F}}}
\newcommand{\Lset}{{\mathcal{L}}}
\newcommand{\Tset}{{\mathcal{T}}}
\newcommand{\Jset}{{\mathcal{J}}}
\newcommand{\neighbourhood}[1]{{\mathcal{N}(#1)}}

\newcommand{\stopset}{{S}}

\newcommand\myor{\hspace{4pt}\kern1pt\rule[-\dp\strutbox]{1pt}{\baselineskip}\kern1pt\kern1pt\rule[-\dp\strutbox]{1pt}{\baselineskip}\kern1pt\hspace{4pt}}

%% file: appendix_A.tex
\label{sec:tbtr_preprocessing}
The TBTR algorithm by \cite{witt2015trip} has a two-stage preprocessing phase. The first stage (Algorithm \ref{alg:algo1}) involves creation of a \textit{trip-transfers set} $\Tset$. This set may contain a large collection of trip-transfers, most of which may never be part of optimal journeys for any source-destination pair. Hence, the second stage (Algorithms \ref{alg:algo2} and \ref{alg:algo3}) aims to reduce $\Tset$ by removing trip-transfers that are guaranteed to never be a part of an optimal journey. These algorithms are reviewed here to keep this paper self-contained. More details including an explanation of the correctness of the algorithms can be found in the original paper. 

\begin{figure}[h]
    \centerline{\includegraphics[scale=0.4]{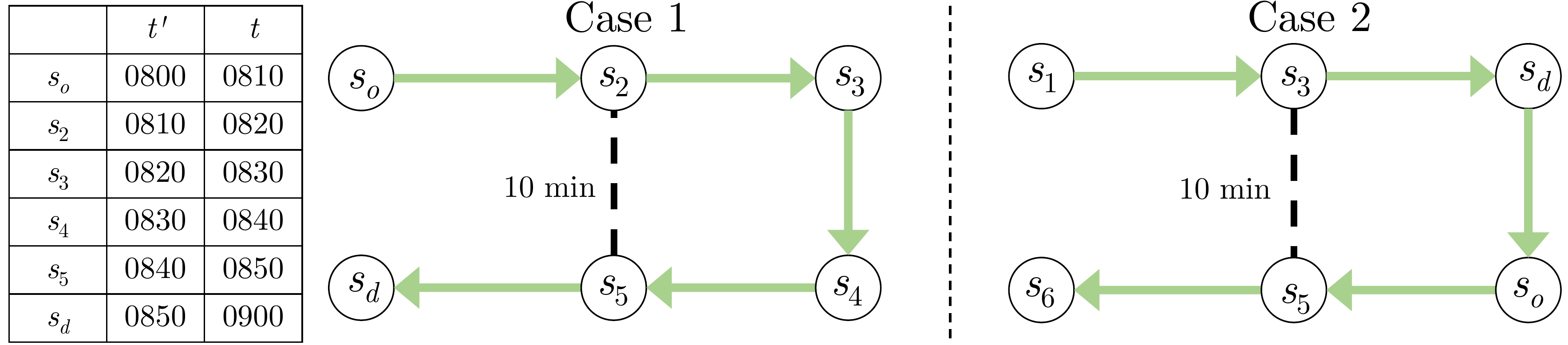}}
    \caption{Special cases for including trip-transfers in Algorithm \ref{alg:algo1}}
    \label{fig:algo1}
    \end{figure}

\textbf{Algorithm \ref{alg:algo1}:} This algorithm generates an initial set of trip-transfers $\Tset$ from the GTFS data by iterating over all stops starting from the second stop of each trip. For each such stop, a stop $s$ in its neighborhood is picked (Line 4). Next, for each route $r$ passing through $s$ (with $s$ not being the last stop on $r$), we find the earliest trip $t^\prime$ on $r$ that the passenger can board, i.e., 
$\triparr{t}{i}+\footpath{\tripstop{t}{i}}{s}\leq \triparr{t^\prime}{j}$, where $i$ and $j$ are the $i^{th}$ and $j^{th}$ stops on trips $t$ and $t'$, respectively. If such a trip exists, the corresponding trip-transfer will be feasible if both the trips are on different routes, i.e., $r\neq\triproute{t}$. Furthermore, transfers between trips on the same route are needed only if  $t^\prime\prec t$ or if $j$ comes before $i$ $(j<i)$. To understand this condition, consider the example in Figure \ref{fig:algo1}. 
Note that the stop IDs of the nodes are different in the subfigures. For Case 1, let $t^\prime$ and $t$ be two trips on the green route such that $t^\prime\prec t$. The arrival times associated with the two trips are shown in the table. Also let $\footstops{s_2}{s_5}\in F$ with $\footpath{s_2}{s_5}=10$ minutes. Thus, for a bicriterion query with $\dep=0805$ and $\source$ and $\target$ as shown, the trip-transfer $\triptrans{t}{2}{t^\prime}{5}$ is a part of the optimal journey. Case 2 illustrates the $j<i$ scenario, i.e., when the passenger transfers to an earlier stop on the same route. Transfers to later trips on the same route or later stops on the same trip are not needed.

\begin{algorithm}[h]
\caption{TBTR: Initial trip-transfer computation}
\label{alg:algo1}
 \hspace*{\algorithmicindent} \textbf{Input}: GTFS\\
 \hspace*{\algorithmicindent} \textbf{Output}: $\Tset$
	\begin{algorithmic}[1]
	\State{$\Tset\leftarrow\emptyset$}
   	\For{$t \in T$}
   	\For{$i = 2, \ldots, \triplen{t}$}
   	\For{ stop $s$ in $\neighbourhood{\tripstop{t}{i}}$}
   	\For{ $(r,j)$ s.t. $r(j)=s \AND j< \routelen{r}$} 

	\State{$t^\prime\leftarrow$ earliest trip on $r$ s.t. $ \triparr{t}{i}+\footpath{\tripstop{t}{i}}{s}\leq \triparr{t^\prime}{j}$}
	\If{$ r \neq \triproute{t}$}
	\State{$\Tset \leftarrow  \Tset \cup \{ \triptrans{t}{i}{t^\prime}{j} \}$}
	\ElsIf{$t^\prime\prec t \OR  j < i$}
	\State{$\Tset \leftarrow  \Tset \cup \{ \triptrans{t}{i}{t^\prime}{j} \}$}
		\EndIf        
        \EndFor
        \EndFor
        \EndFor
        \EndFor
	\end{algorithmic}
\end{algorithm}
    
\begin{figure}[h]
    \centerline{\includegraphics[scale=0.45]{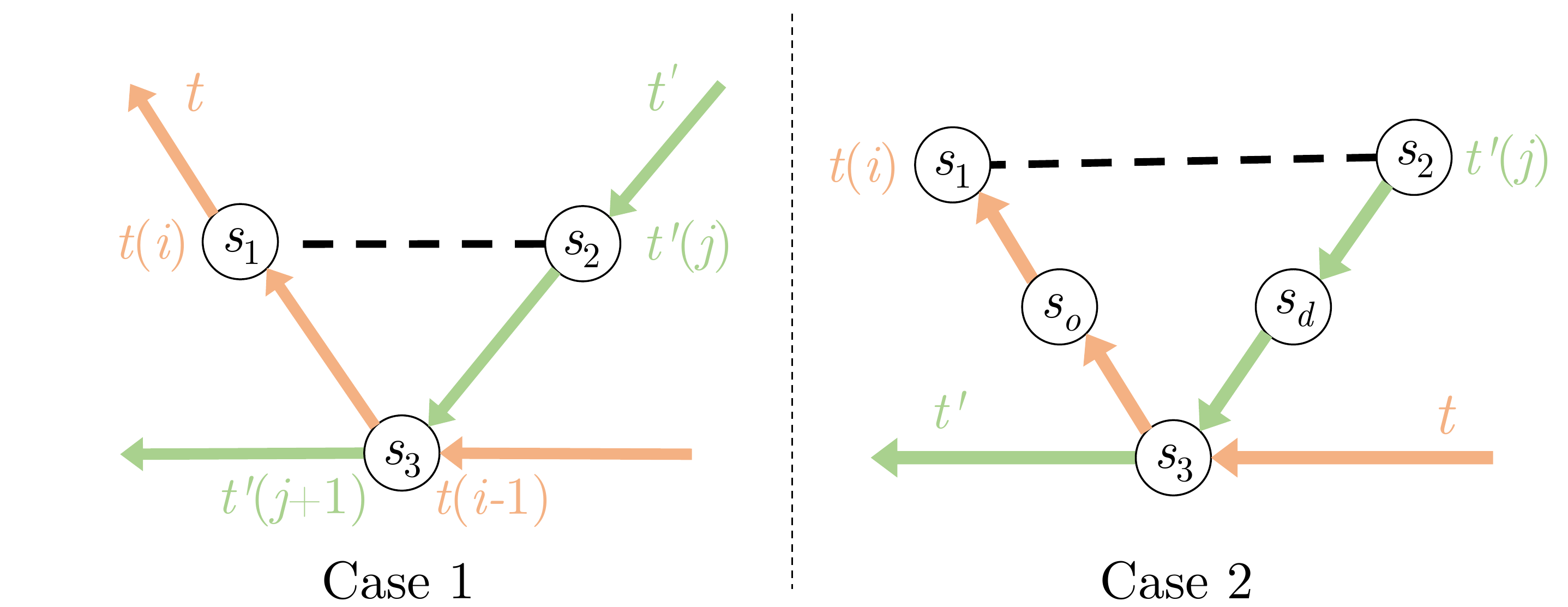}}
    \caption{Removing U-turn like movements using Algorithm \ref{alg:algo2}}
    \label{fig:Utrun}
    \end{figure}

\textbf{Algorithm \ref{alg:algo2}:} This algorithm uses the GTFS data and the output from Algorithm \ref{alg:algo1} to reduce the size of the trip-transfer set $\Tset$. The objective here is to remove trip-transfers from $\Tset$ that result in a U-turn like movement. For illustration see Case 1 of Figure \ref{fig:Utrun}. The example shown considers two trips $t$ and $t^\prime$ for which $\tripstop{t}{i}=s_1$, $\tripstop{t^\prime}{j}=s_2$, and $\tripstop{t}{i-1}=\tripstop{t^\prime}{j+1}=s_3$. Further, let $\footstops{s_1}{s_2}\in F$ be such that $\triptrans{t}{i}{t^\prime}{j}\in \Tset$.  According to Line 2 of the algorithm, if $\triparr{t}{i-1}\leq\tripdep{t'}{j+1}$, trip-transfer $\triptrans{t}{i}{t^\prime}{j}$ is never part of an optimal journey because a passenger traveling on $t$ can switch to $t^\prime$ at stop $s_3$. To understand the need for condition $\tripstop{t}{i-1}=\tripstop{t}{j+1}$ while checking for U-turns, consider Case 2. Here, the trip-transfer $\triptrans{t}{i}{t^\prime}{j}$ may be part of an optimal journey between $\source$ and $\target$ and hence, cannot be removed. 

\begin{algorithm}[h]
\caption{TBTR: Remove trip-transfers resulting in U-turns}
\label{alg:algo2}
 \hspace*{\algorithmicindent} \textbf{Input}: GTFS, $\Tset$\\
 \hspace*{\algorithmicindent} \textbf{Output}: $\Tset$
	\begin{algorithmic}[1]
	\vspace{2mm}
   	\For{$\triptrans{t}{i}{t^\prime}{j} \in \Tset$}
   	\If{$\tripstop{t}{i-1} = \tripstop{t^\prime}{j+1}$ \textbf{and} $\triparr{t}{i-1}\leq\tripdep{t^\prime}{j+1}$}
	\State{$\Tset \leftarrow \Tset\hspace{1pt}\backslash\hspace{1pt}\left\{ \triptrans{t}{i}{t^\prime}{j}\right\}$}	

		\EndIf        
        \EndFor
	\end{algorithmic}
\end{algorithm}

\textbf{Algorithm \ref{alg:algo3}:} The set $\Tset$ can be further reduced to only contain trip-transfers that result in improved arrival times at any stop. To this end, each trip $t$ is scanned in the decreasing order of its stop sequence. For every trip $t$, we maintain an arrival time label $\prearr{s}$ for each stop $s \in \stopset$ (Lines 1--2). The $\prearr{s}$ variable need not be stored for each trip and can be overwritten. Hence, we skip the trip index in its notation.

\begin{algorithm}[H]
\caption{TBTR: Remove non-optimal trip-transfers}
\label{alg:algo3}
 \hspace*{\algorithmicindent} \textbf{Input}: GTFS, $\Tset$\\
 \hspace*{\algorithmicindent} \textbf{Output}: $\Tset$
	\begin{algorithmic}[1]
	\For{$t\in T$}
	\State{$\prearr{s}\leftarrow \infty \hspace{6pt}\forall \hspace{2pt}s\in S$}
	\For{$i = \triplen{t}, \ldots, 2$}
	\For{$s \in\neighbourhood{\tripstop{t}{i}}$}	
	\State{$\prearr{s}\leftarrow \min \big\{\prearr{s}, \triparr{t}{i}+ \footpath{t(i)}{s} \big\}$}
    \EndFor
	\For{$\triptrans{t}{i}{t^\prime}{j}\in \Tset$}
	\State{$keep \leftarrow \textbf{false}$}
	\For{ stop $\tripstop{t^\prime}{k}$ with $k>j$}
	\For{$s \in \neighbourhood{\tripstop{t^\prime}{k}}$}
	\If{$\triparr{t^\prime}{k}+ \footpath{\tripstop{t^\prime}{k}}{s} < \prearr{s}$}
	\State{$keep \leftarrow\textbf{true}$ }
	\State{$\prearr{s}\leftarrow \triparr{t^\prime}{k}+ \footpath{\tripstop{t^\prime}{k}}{s}$}
	\EndIf
    \EndFor
    \EndFor

	\If{ \textbf{not} $keep$}
	\State{$\Tset \leftarrow \Tset\hspace{1pt}\backslash\hspace{1pt}\left\{ \triptrans{t}{i}{t^\prime}{j}\right\}$}

		\EndIf
        \EndFor
        \EndFor
        \EndFor
	\end{algorithmic}
\end{algorithm}

While iterating over the $i^{th}$ stop of trip $t$, we first improve the arrival time labels of all the stops in the neighborhood of stop $\tripstop{t}{i}$ if possible (Lines 3--5). Next, for each trip-transfer $\triptrans{t}{i}{t^\prime}{j}$ from the stop $\tripstop{t}{i}$, we check if taking the transfer improves the arrival time at any stop (Lines 6--12). (The symbols $t$ and $i$ in Line 6 are assumed to be the same as the trip $t$ and stop index $i$ from Lines 1 and 3.) This is done by iterating over all the stops of $t^\prime$ from $j$ onwards and by performing a similar update on the $\prearr{.}$ labels. The trip-transfer is discarded if it does not result in an improved label (Lines 13--14). \cite{lehoux_et_al:OASIcs:2020:13139} proposed an additional route-based pruning scheme that accelerated  trip-transfer computations. However, unlike several other PTR algorithms, their methods were tested on instances with footpath graphs that were not necessarily transitively closed. 

Table \ref{tab:tbtrpre} summarizes statistics associated with the above mentioned algorithms on the six test cases. For each algorithm, two metrics are reported---\textit{$\Tset$-time}$^*$ and \textit{$\Tset$-size}. \textit{$\Tset$-time}$^*$ is the time (in seconds) required to compute and refine $\Tset$ and \textit{$\Tset$-size} is the number of trip-transfers in $\Tset$ (in millions) at the end of preprocessing. 

\begin{table}[H]
    \caption{TBTR preprocessing statistics (\textit{$\Tset$-size}: size of $\Tset$ in millions, \textit{$\Tset$-time}$^*$: time in seconds to find $\Tset$)}
    \begin{center}
    \begin{tabular}{c g c g c }
    \hline
    \textbf{Network} &
       \textbf{Metric} & \textbf{Algorithm  \ref{alg:algo1}} &
       \textbf{Algorithm  \ref{alg:algo2}}&
       \textbf{Algorithm  \ref{alg:algo3}}\\
    \hline
    \multirow{2}{*}{Switzerland} & \textit{$\Tset$-size} & 20.2 & 17.6 & 2.4 \\
     & \textit{$\Tset$-time$^*$} & 32.3 & 50.3 & 40.3\\
    \cdashline{1-5}
    \multirow{2}{*}{Netherlands} & \textit{$\Tset$-size }& 9.9 & 9.8 & 2.1 \\
     & \textit{$\Tset$-time$^*$} & 18& 36.8 & 15.3\\
    \cdashline{1-5}
    \multirow{2}{*}{Sweden} & \textit{$\Tset$-size} & 26.2 & 23.9 & 0.7\\
     & \textit{$\Tset$-time$^*$} & 27.7 & 132 & 420.2\\
    \cdashline{1-5}
    \multirow{2}{*}{Israel} & \textit{$\Tset$-size } & 76.7 & 76.0 & 7.5 \\
     & \textit{$\Tset$-time$^*$} & 205.2 & 271.6 & 225.5\\
    \cdashline{1-5}
    \multirow{2}{*}{Taichung} & \textit{$\Tset$-size }&  9.3& 8.7 & 1.4 \\
     & \textit{$\Tset$-time$^*$} & 41.3 & 28.1 & 17.5 \\
    \cdashline{1-5}
    \multirow{2}{*}{Bangalore} & \textit{$\Tset$-size }&  230.8 & 228.3 & 14.1  \\
     & \textit{$\Tset$-time$^*$} & 235.1 & 849.9 & 878.1\\
    \hline
\end{tabular}
\label{tab:tbtrpre}
\end{center}
\end{table}

%% file: appendix_B.tex
\label{sec:raptor_appen}
Algorithm \ref{alg:dp} presents the pseudocode for the standard bicriterion RAPTOR algorithm \citep{delling2015round}. The list of stops whose labels improved in the previous round is indicated by \textit{mlist} and is initialized with the source $\source$. 

\begin{algorithm}[H]
\caption{RAPTOR: Bicriterion query}
\label{alg:dp}
 \hspace*{\algorithmicindent} 
    \textbf{Input}: GTFS, $\source$,  $\target$, $\dep$, $\maxtrans$\\
 \hspace*{\algorithmicindent} 
    \textbf{Output}: Optimal labels 
	\begin{algorithmic}[1]
	\vspace{2mm}
		\State $\tau_{opt}(n, s), \tau^*(s) \leftarrow\infty, \infty ~  \forall~s\in S\backslash\{\source\},\hspace{2pt} \forall \hspace{2pt} n\leq \maxtrans$ 
		\State $\tau_{opt}(0,\source), \tau^*(\source)\leftarrow\dep, \dep$
		\State $mlist\leftarrow \big\{\source\big\}$
		\vspace{2mm}
		
		\For{$n = 1, 2, \ldots, \maxtrans$} 
		\State $Q\leftarrow\emptyset$ 
        \For{$s\in mlist$} 
        	\For{routes $r$ s.t. $r(i)=s$ for some index $i \in \{1,\ldots,\routelen{r}\}$}
    		\If{$(r, r(j)) \in Q$ for some index $j \in \{0,\ldots,\routelen{r}\}$}
    		    \State replace $(r,r(j))$ in $Q$ by $(r,r(i))$ if $i<j$
    		\Else{ $Q\leftarrow Q\cup\big\{(r,r(i))\big\}$}
    		\EndIf
    	    \EndFor				
        \State $mlist\leftarrow mlist\backslash\{s\}$
    	    \EndFor				
    	   \vspace{2mm}

    	 \For{$(r,r(i))\in Q$}
    	 \State  $t\leftarrow null$ 
    	    \For{$j = i,\ldots,\routelen{r}$}
    	    \If{$t\neq null~\AND~\triparr{t}{j} <\min\big\{\tau^*(r(j)),$ $\tau^*(\target)\big\}$}
    	    \State{$\tau_{opt}(n,r(j)) \leftarrow \triparr{t}{j}$}
    	    \State{$\tau^*(r(j)) \leftarrow \triparr{t}{j}$}
    	    \State $mlist \leftarrow mlist\cup \big\{r(j)\big\}$
            \EndIf
    	    \If {$\tau_{opt}(n-1, r(j)) < \tripdep{t}{j}$}
    	    \State $t\leftarrow$ earliest trip on $r$ from stop $r(j)$ 
            \EndIf
    	    \EndFor				
    	 \EndFor
    	 \vspace{2mm}

    	 \For{$s\in mlist$}
    	    \For{$s^\prime \in \neighbourhood{s}$}
            \If{$\tau_{opt}(n, s)+\footpath{s}{s^\prime}<\tau^*(s^\prime)$}
    	    \State $\tau_{opt}(n,s^\prime)\leftarrow \tau_{opt}(n,s)+\footpath{s}{s^\prime}$
    	    \State{$\tau^*(s^\prime) \leftarrow \tau_{opt}(n,s)+\footpath{s}{s^\prime}$}
    	    \State $mlist \leftarrow mlist\cup \big\{s^\prime\big\}$
        \EndIf
        \EndFor
        \EndFor
        \vspace{2mm}

        \If{$mlist$ is empty}
        \State{\textbf{break}}
        \EndIf
        \EndFor				
	\end{algorithmic}
\end{algorithm}
\cite{delling2015round} also modified RAPTOR to rRAPTOR which handles range queries. Building on rRAPTOR, Algorithm \ref{alg:One-To-Many_rRAPTOR} outlines the pseudocode for our version of One-To-Many rRAPTOR. Given a list of destinations to which optimal journeys are sought, we update a stop label only if the new arrival time at the stop is less than the maximum of the labels of stops in the destination list. To do so, in each round, Line 10 initializes $\tau^*_{max}$ to the maximum of destination labels. Pruning conditions in Lines 22 and 32 are verified as described above. The algorithm also prunes the destination list using Lines 7--9 motivated by the ideas discussed in Section \ref{sec:OneToMany}. Specifically, for every departure time, we initialize $dlist^\prime$ to a copy of  $dlist$ in Line 5. Next, we find $\tau^*_{min}$, the minimum of labels updated in the previous round. If the best arrival time at $s_d\in dlist^\prime$ is less than $\tau^*_{min}$, we can safely remove it from $dlist^\prime$ since we cannot improve the labels of $s_d$ as all updates in the current(later) round(s) will be greater than $\tau^*_{min}$ (since Pareto-optimality requires that the arrival time should decrease with the number of rounds).

\begin{algorithm}[H]
\caption{One-To-Many rRAPTOR}
\label{alg:One-To-Many_rRAPTOR}
 \hspace*{\algorithmicindent} 
    \textbf{Input}: GTFS, $\source$, $\maxtrans$, $dlist$, $tlist$\\
 \hspace*{\algorithmicindent} 
    \textbf{Output}: Optimal labels
	\begin{algorithmic}[1]
	\vspace{2mm}
	\State $\tau_{opt}(n,s), \tau^*(s) \leftarrow\infty, \infty\  \forall~s\in S\backslash\{\source\},\hspace{2pt} \forall \hspace{2pt} n\leq \maxtrans$ 
    \For{$\dep \in tlist$}
		\State $\tau_{opt}(0,\source), \tau^*(\source)\leftarrow \dep, \dep$
		\State $mlist\leftarrow \big\{\source\big\}$
        \State $dlist^\prime\leftarrow dlist$

		\vspace{2mm}
		\For{$n = 1, 2, \ldots, \maxtrans$} 
		\State $\tau^*_{min}\leftarrow \min_{s \in mlist} \tau^*(s)$ 
	    \For{$s\in dlist^\prime$ s.t. $\tau^*_{min}\leq\tau^*(s)<\infty$}
        \State $dlist^\prime\leftarrow dlist^\prime\backslash\big\{s\big\}$
        \EndFor				
		\State $\tau^*_{max}\leftarrow \max_{s \in dlist^\prime}\tau^*(s)$

		\State $Q\leftarrow\emptyset$ 
        \For{$s\in mlist$} 
        	\For{routes $r$ s.t. $r(i)=s$ for some index $i \in \{0,\ldots,\routelen{r}\}$}
    		\If{$(r, r(j)) \in Q$ for some index $j \in \{0,\ldots,\routelen{r}\}$}
    		    \If{$i<j$}
    		    \State replace $(r,r(j))$ in $Q$ by $(r,r(i))$ 
        		\EndIf
    		\Else{ $Q\leftarrow Q\cup\big\{(r,r(i))\big\}$}
    		\EndIf
    	    \EndFor				
        \State $mlist\leftarrow mlist\backslash\big\{s\big\}$
    	    \EndFor				
    	   \vspace{2mm}

    	 \For{$(r,r(i))\in Q$}
    	 \State  $t\leftarrow null$ 
    	    \For{$j = i,\ldots,\routelen{r}$}
    	    \If{$t\neq null~\AND~\triparr{t}{j} < \min\big\{\tau^*(r(j)), \tau^*_{max}\big\}$}
    	    \State{$\tau_{opt}(n,r(j)) \leftarrow \triparr{t}{j}$}
    	    \State{$\tau^*(r(j)) \leftarrow \triparr{t}{j}$}
    	    \State $mlist \leftarrow mlist\cup \big\{r(j)\big\}$
    	    \If{$r(j) \in dlist^\prime$}
    		\State $\tau^*_{max}\leftarrow \max_{s \in dlist^\prime}  \tau^*(s)$
            \EndIf
            \EndIf
    	    \If {$\tau_{opt}(n-1,r(j)) < \tripdep{t}{j}$}
    	    \State $t\leftarrow$ earliest trip on $r$ from stop $r(j)$ 
            \EndIf
    	    \EndFor				
    	 \EndFor
    	 \vspace{2mm}

    	 \For{$s\in mlist$}
    	    \For{$s^\prime \in \neighbourhood{s}$}
            \If{$\tau_{opt}(n,s)+\footpath{s}{s^\prime}<\min\big\{ \tau^*(s^\prime),\tau^*_{max}\big\}$}
    	    \State $\tau_{opt}(n,s^\prime)\leftarrow \tau_{opt}(n,s)+\footpath{s}{s^\prime}$
    	    \State{$\tau^*(s^\prime) \leftarrow \tau_{opt}(n,s)+\footpath{s}{s^\prime}$}
    	    \State $mlist \leftarrow mlist\cup \big\{s^\prime\big\}$
    	    \If{$s^\prime \in dlist^\prime$}
    		\State $\tau^*_{max}\leftarrow \max_{s \in dlist^\prime}  \tau^*(s)$
    \EndIf
        \EndIf
        \EndFor
        \EndFor
        \vspace{2mm}

        \If{$mlist$ is empty}
        \State{\textbf{break}}
        \EndIf
        \EndFor				
    \EndFor
	\end{algorithmic}
\end{algorithm}

%% file: appendix_C.tex
\label{sec:weighting_scheme_compari}
In this section, we study the effect of different weighing schemes discussed in Section \ref{subsec:HypTBTR}. The partitioning algorithm used is KaHyPar and the experimental setup and metrics presented are same as that in Section \ref{sec:Experiments}. Tables \ref{tab:sc1__sc2_scheme} summarizes the performance metrics of the Sc$_1$ (unweighted) and Sc$_2$ scheme. These results are compared with the Sc$_3$ from Table \ref{tab:hyptbtr_pre_results}.

\begin{table}[h]
    \caption{HypTBTR preprocessing using KaHyPar with Sc$_1$ and Sc$_2$ scheme (\textit{scut}: cutstop count and \% of cutstops, \textit{pqueries}: profile queries required, \textit{$\fillintrip$ size}: \% of fill-in trips, \textit{$\fillintrip$ time}$^*$: time in seconds to compute $\fillintrip$. Values in teal indicate \% gain in the multilevel version over its standard counterpart.)\vspace{-5mm}}
     \begin{center}
    \begin{tabular}{c c c g c g c}
    \hline
    \multirow{2}{*}{\textbf{Network}} & \multirow{2}{*}{\textbf{Partitioning}}&
    \multirow{2}{*}{\textbf{Metric}} & \multicolumn{4}{c}{\textbf{Partitions}}\\
    &&& \textbf{4 (2-2)}& \textbf{6 (3-2)}& \textbf{8 (4-2)}& \textbf{10 (5-2)}\\
    \hline
    \multirow{8}{*}{\makecell{Sweden\\(Sc$_1$)}} & \multirow{4}{*}{standard}& \textit{scut}     &   344 (1\%)    &     347 (1\%)    &    362 (1\%)&   362 (1\%)\\
    & &\textit{pqueries} &  117\smp992    & 120\smp062   & 130\smp682 & 130\smp682 \\
    & &\textit{$\fillintrip$ size}   & 0.6\%      & 0.7\%    & 1.3\% & 1.4\%\\
    & &\textit{$\fillintrip$ time}$^*$   &  538.1    &     526.5 & 542.3 & 535.8\\[.15cm]
    \cdashline{3-7}
     \rule{0pt}{.45cm} & \multirow{4}{*}{multilevel} & \textit{scut}     &   191 (0.6\%)    &     360 (1\%)    &    364 (1\%)&   362 (1\%)\\
    & &\textit{pqueries} &   36\smp157   \textcolor{teal}{(69.4\%)} & 129\smp276   \textcolor{teal}{(-7.7\%)}& 131\smp844 \textcolor{teal}{(-0.9\%)}& 118\smp335 \textcolor{teal}{(9.4\%)}\\
    & &\textit{$\fillintrip$ size}   & 0.6\%      & 0.6\%    & 0.6\% & 0.6\%\\
    & &\textit{$\fillintrip$ time}$^*$   &  157.8 \textcolor{teal}{(70.7\%)}    &    510.5 \textcolor{teal}{(3.0\%)} & 550.1 \textcolor{teal}{(-1.4\%)} & 530.6 \textcolor{teal}{(1.0\%)}\\
    \hline
    \multirow{8}{*}{\makecell{Sweden\\(Sc$_2$)}} & \multirow{4}{*}{standard}& \textit{scut}     &   78 (0.2\%)    &     141 (0.4\%)    &    167 (0.5\%)&   228 (0.7\%)\\
    & &\textit{pqueries} &  6\smp006    &   19\smp740 & 27\smp722 & 51\smp756 \\
    & &\textit{$\fillintrip$ size}   & 6.4\%      & 14\%    & 13.6\% & 18.4\%\\
    & &\textit{$\fillintrip$ time}$^*$   &    10.2  & 15.8    & 18.1 & 23.2\\[.15cm]
    \cdashline{3-7}
     \rule{0pt}{.45cm} & \multirow{4}{*}{multilevel} & \textit{scut}     &   95 (0.3\%)    &     135 (0.4\%)    &    179 (0.5\%)&   225 (0.6\%)\\
    & &\textit{pqueries} &  3\smp096  \textcolor{teal}{(48.5\%)}  &   8\smp086 \textcolor{teal}{(59\%)} & 13\smp582 \textcolor{teal}{(51\%)} & 21\smp572 \textcolor{teal}{(58.3\%)}\\
    & &\textit{$\fillintrip$ size}   & 2.5\%      & 9.1\%    & 11.4\% & 16.1\%\\
    & &\textit{$\fillintrip$ time}$^*$   &      5.7 \textcolor{teal}{(44.1\%)}   & 8.5    \textcolor{teal}{(46.2\%)} & 9.3 \textcolor{teal}{(48.6\%)} & 12.9 \textcolor{teal}{(44.4\%)}\\
    \hline

    \end{tabular}
\label{tab:sc1__sc2_scheme}
    \end{center}
\end{table}

Since the nodes and edges are unweighted in Sc$_1$, KaHyPar's objective reduces to minimizing the number of cut hyperedges. Hence, the size of the route cells varies greatly. The size of a route cell is defined as the sum of events associated with stops belonging to the route cell. For example, in Sweden, using standard partitioning, the ratio of the largest route cell size to the smallest varies in the range of 31--823. For Sc$_3$ and Sc$_2$, this ratio varies between 1--5. Recall that HypTBTR (and HypRAPTOR) only scans trips belonging to the source/destination route cells and the fill-in during the query phase. Thus, the greater the differences in the sizes of partitions generated, the larger is the variation in query times. Furthermore, we anticipate the fill-in computation time to increase with the number of partitions (due to increase in cutstops). While Sc$_2$ and Sc$_3$ show the expected trend, the time required to compute the fill-in for Sc$_1$ was found to be erratic. In terms of \textit{scut}, $\fillintrip$ $size$, and $\fillintrip$ $time^*$, Sc$_3$ was found to marginally outperform Sc$_2$.

%% file: appendix_D.tex
\label{sec:hmetis_vs_kahypar}
In Section \ref{subsec:HypTBTR}, three methods for hypergraph partitioning were discussed: KaHyPar, hMETIS, and an IP. Table \ref{tab:hmetis_table} shows the performance of the HypTBTR algorithm on the Sweden network using hMETIS with the following parameters: \textit{UBfactor} 15, \textit{Nruns} 50, \textit{Ctype} 1, \textit{Rtype} 1, \textit{Vcycle} 1, \textit{Reconst} 0, and \textit{dbglvl} 0. The metrics presented are same as those discussed in Section \ref{sec:Experiments}.

Comparing the number of cutstops and the fill-in size with the corresponding values in Table \ref{tab:hyptbtr_pre_results}, we see that KaHyPar performs marginally better than hMETIS. A drawback of the hMETIS implementation \href{http://glaros.dtc.umn.edu/gkhome/metis/hmetis/download}{(hMETIS Code)} used is that the seed could not be fixed and hence it was not possible to replicate results in repeated runs. We did not face this issue with KaHyPar since it allows configuring the seed.
\begin{table}[h]
    \caption{HypTBTR preprocessing using hMETIS (\textit{scut}: cutstop count and \% of cutstops, \textit{pqueries}: profile queries required, \textit{$\fillintrip$ size}: \% of fill-in trips, \textit{$\fillintrip$ time}$^*$: time in seconds to compute $\fillintrip$. Values in teal indicate \% gain in the multilevel version over its standard counterpart.)\vspace{-5mm}}
     \begin{center}
    \begin{tabular}{c c c g c g c}
    \hline
    \multirow{2}{*}{\textbf{Network}} & \multirow{2}{*}{\textbf{Partitioning}}&
    \multirow{2}{*}{\textbf{Metric}} & \multicolumn{4}{c}{\textbf{Partitions}}\\
    &&& \textbf{4 (2-2)}& \textbf{6 (3-2)}& \textbf{8 (4-2)}& \textbf{10 (5-2)}\\
    \hline
    \multirow{8}{*}{Sweden} & \multirow{4}{*}{standard}& \textit{scut}     & 93  (0.3\%)    &   139  (0.4\%)    &    213 (0.6\%)&   223 (0.6\%)\\
    & &\textit{pqueries} &   8\smp556   &  19\smp182  &  45\smp156 &  49\smp506\\
    & &\textit{$\fillintrip$ size}   & 6.9\%      & 11.8\%    & 16.6\% & 16.4\%\\
    & &\textit{$\fillintrip$ time}$^*$   &   10.4   & 11.8    & 18.7 & 24.1\\[.15cm]
    \cdashline{3-7}
     \rule{0pt}{.45cm} & \multirow{4}{*}{multilevel} & \textit{scut}     &  62 (0.2\%)    &    134 (0.4\%)    &   195 (0.6\%)&  208 (0.6\%)\\
    & &\textit{pqueries} &   1\smp995 \textcolor{teal}{(76.7\%)}   &  6\smp564 \textcolor{teal}{(65.8\%)}  & 18\smp723 \textcolor{teal}{(58.5\%)} & 20\smp712 \textcolor{teal}{(58.2\%)} \\
    & &\textit{$\fillintrip$ size}   & 1.9\%      & 10.2\%    & 12.6\% & 13.5\%\\
    & &\textit{$\fillintrip$ time}$^*$   & 3.5 \textcolor{teal}{(66.3\%)}     & 6.5 \textcolor{teal}{(44.9\%)}    & 9.1 \textcolor{teal}{(51.3\%)} & 10.3 \textcolor{teal}{(57.3\%)}\\
    \hline
    \end{tabular}
\label{tab:hmetis_table}
    \end{center}
\end{table}

Results from partitions generated by the IP formulation are presented in Table \ref{tab:ILP}. The optimization model was solved using CPLEX. While implementations of KaHyPar \href{https://github.com/kahypar/kahypar}{(github.com/kahypar)} and hMETIS \href{http://glaros.dtc.umn.edu/gkhome/metis/hmetis/download}{(hMETIS Code)} can generate partitions in a few seconds, the IP model could take several hours to converge to the optimal solution, especially for large networks. E.g., for partitioning the Sweden network into two partitions, CPLEX took about 15 minutes to achieve a gap of 0.01\%. However, for four partitions, even after four days, the gap reduced to only 1.01\%. Because of the longer run times, we only report results for four partitions and the corresponding 2-2 multilevel case. Note that there is a significant drop in the values of \textit{scut} and \textit{$\fillintrip$ size} from the standard to the multilevel version. This could possibly be due to the higher optimality gap for the 4-way standard partitioning. 

\begin{table}[h]
    \caption{HypTBTR preprocessing using IP (\textit{scut}: cutstop count and \% of cutstops, \textit{pqueries}: profile queries required, \textit{$\fillintrip$ size}: \% of fill-in trips, \textit{$\fillintrip$ time}$^*$: time in seconds to compute $\fillintrip$. Values in teal indicate \% gain in the multilevel version over its standard counterpart.) \vspace{-5mm}}
     \begin{center}
    \begin{tabular}{c c c g}
    \hline
    \multirow{2}{*}{\textbf{Network}} & \multirow{2}{*}{\textbf{Partitioning}}&
    \multirow{2}{*}{\textbf{Metric}} & \multicolumn{1}{c}{\textbf{Partitions}}\\
    &&& \textbf{4 (2-2)}\\
    \hline
    \multirow{8}{*}{Sweden} & \multirow{4}{*}{standard}& \textit{scut}     &  273 (0.8\%)    \\
    & &\textit{pqueries} &   74256   \\
    & &\textit{$\fillintrip$ size}   & 13.6\%      \\
    & &\textit{$\fillintrip$ time}$^*$   & 19.3  \\[.15cm]
    \cdashline{3-4}
     \rule{0pt}{.45cm} & \multirow{4}{*}{multilevel} & \textit{scut}     & 49  (0.1\%)  \\
    & &\textit{pqueries} &      1422\textcolor{teal}{(98.1\%)} \\
    & &\textit{$\fillintrip$ size}   & 1.7\%      \\
    & &\textit{$\fillintrip$ time}$^*$   &     3.2 \textcolor{teal}{(83.4\%)}\\
    \hline
    \end{tabular}
\label{tab:ILP}
    \end{center}
\end{table}